\documentclass[aps,prb,preprint,nofootinbib,twocolumn,10pt]{revtex4-2}

\usepackage{hyperref}
\hypersetup{
    colorlinks=true,
    citecolor=blue,
    linkcolor=blue,
    filecolor=magenta,      
    urlcolor=blue,
    pdftitle={Overleaf Example},
    pdfpagemode=FullScreen,
    }

\usepackage{cprotect} 
\usepackage{listings}
\usepackage{graphicx}
\usepackage{dcolumn}
\usepackage{bm}
\usepackage{amsmath}
\usepackage{multirow}
\usepackage{xcolor}
\def\*#1{\mathbf{#1}}

\begin{document}
\title{Convergence of body-orders in linear atomic cluster expansions}
\author{Apolinario Miguel Tan}
    \email{atan@sissa.it}
\affiliation{Scuola Internazionale Superiore di Studi Avanzati, Trieste, Italy}
\author{Franco Pellegrini}
\affiliation{Scuola Internazionale Superiore di Studi Avanzati, Trieste, Italy}
\author{Stefano de Gironcoli}
\affiliation{Scuola Internazionale Superiore di Studi Avanzati, Trieste, Italy}

\date{\today}
\begin{abstract}
  We study the convergence of a linear atomic cluster expansion (ACE) potential \cite{drautzAtomicClusterExpansion2019} with respect to its basis functions, in terms of the effective two-body interactions of elemental Carbon and Silicon systems. We build ACE potentials with descriptor sets truncated at body-orders $K=2$ to $K=5$ trained on a diverse Carbon dataset and on Silicon dimers to pentamers. The potentials trained on diverse structures with standard ACE bases are not able to recover the correct dimer curves much less produce stable curves or solutions. While employing ACE bases removed of self-interactions still does not generalize to the DFT-expected results, properly tailored datasets and basis sets are able to show signs of convergence and stability in the curves and expansions, suggesting a method to build potentials with interpretable bases with respect to the cluster expansion. 
\end{abstract}
\maketitle
\section{Introduction}
Decompositions of the total energy of atomic systems by $N$-body contributions, or cluster expansions,
\begin{align}
    E\left(\{\*r_i\}\right) &= V^{(0)} + \sum_i V^{(1)}\left(\*r_i\right) + \sum_{\substack{i,j\\i\neq j}}V^{(2)}\left(\*r_i,\*r_j\right) + \notag \\ & \sum_{\substack{i,j,k\\i\neq j\neq k\neq i}}V^{(3)}\left(\*r_i,\*r_j,\*r_k\right) + ...
    \label{eq:cluster_expansion}
\end{align}
have been used as the framework to study material properties such as the thermodynamics of substitutional alloys \cite{sanchezGeneralizedClusterDescription1984,ferreiraFirstprinciplesCalculationAlloy1989a} and empirical potentials of condensed matter and molecular systems \cite{stillingerComputerSimulationLocal1985,biswasNewClassicalModels1987,tersoffModelingSolidstateChemistry1989}. While the expansion has been shown to be complete \cite{sanchezGeneralizedClusterDescription1984,ferreiraFirstprinciplesCalculationAlloy1989a,dussonAtomicClusterExpansion2022} when the number of atoms matches the highest body order of the expansion, its practical effectiveness is limited to cases where the total energy converges after a few body-orders $\nu$\cite{stillingerComputerSimulationLocal1985,biswasNewClassicalModels1987,drautzGeneralRelationsManybody2004}, as the number of terms increases combinatorially as higher orders are included. 

More recently, the expansion has been used in describing local chemical environments for machine learning potentials. The underlying assumption for these descriptors is that the energy of the system is a sum of its individual atomic contributions $E_i$ \cite{behlerGeneralizedNeuralNetworkRepresentation2007a},
\begin{align}
    E\left(\{\*r_i\}\right) &= \sum_{i=1}^{N} E_i\left(\{\*r_i\}\right)
    \label{eq:atomic_energy_sum}
\end{align}
where $N$ indicates the total number of atoms in the structure. The atomic energies may then be expressed as an expansion in the number of neighbors.

Improvements on the construction of $E_i$ have been made in recent decades, with symmetry-respecting descriptors that use only a few terms in the expansion \cite{behlerGeneralizedNeuralNetworkRepresentation2007a,bartokRepresentingChemicalEnvironments2013,zhangDeepPotentialMolecular2018,lotPANNAPropertiesArtificial2020} or take advantage of tensorial operations to reach an arbitrary body order \cite{shapeevMomentTensorPotentials2016,drautzAtomicClusterExpansion2019,batatiaDesignSpaceE3equivariant2025,NEURIPS2022_4a36c3c5}. A detailed review on descriptors of local chemical environments has been done by \citeauthor{musilPhysicsInspiredStructuralRepresentations2021} \cite{musilPhysicsInspiredStructuralRepresentations2021}

A popular version of this resummation is the Atomic Cluster Expansion proposed by Drautz \cite{drautzAtomicClusterExpansion2019}. In this approach, each $E_i$ is expanded with a non-orthogonal basis, $\{A_{\nu}\}$, with unrestricted sums,
\begin{align}
    E_i\left(\{\*r_i\}\right) &= \sum_\nu c^{(1)}_\nu A_{i,\nu} + \sum_{\nu_1\nu_2}^{\nu_1\geq\nu_2}c_{\nu_1\nu_2}^{(2)}A_{i,\nu_1}A_{i,\nu_2} + ...
    \label{eq:ace}
\end{align}
with single sums over neighbors
\begin{align}
    A_{i,\nu} = \sum_j f_\nu\left(\*r_i,\*r_j\right),
\end{align}
and $\nu$ indexing the terms of the expansion.
The expansion has been shown to be complete \cite{dussonAtomicClusterExpansion2022}, and its implementations are effective in machine learning applications \cite{kovacsLinearAtomicCluster2021,lysogorskiyPerformantImplementationAtomic2021,bochkarevEfficientParametrizationAtomic2022,qamarAtomicClusterExpansion2023,wittACEpotentialsjlJuliaImplementation2023a}. One of the state-of-the-art potentials, MACE \cite{NEURIPS2022_4a36c3c5}, extends the formalism of ACE as a graph neural network but the interpretability of the expansion is traded for better expressivity when compared to linear potentials. 

In employing interatomic potentials using the ACE framework, the selection of the basis complexity in the number of functions and maximum body-order $K$ is a heuristic process and no systematic construction has been proposed yet. While ACE potentials have been shown to achieve low fitting errors, the stability of these solutions and the interpretability of the basis functions are not as studied, beyond undertakings to eliminate the self-interactions in \citet{hoACEwoselfinteraction2024} and perform uncertainty quantifications in linear models in \citet{chongPredictionRigiditiesDatadriven2025}. 

In the advent of foundational machine learning models \cite{batatiaFoundationModelAtomistic2024} as baselines training neural network potentials that do not need to be trained from scratch, it may also be worth asking if a method to systematically build a body-ordered basis with reasonable lower-order energetics can be established. In this work, we investigate the convergence of linear ACE at different body-orders $K$ in terms of their projection on the lowest terms of the original cluster expansion Eq.~\ref{eq:cluster_expansion}. In particular, we consider the convergence of the potential in energy and forces on different datasets as a function of the number and type of basis functions utilized, and compare it to the stability of a real body-ordered expansion. Our study focuses on the first nontrivial term of the expansion: the two-body term. This can be easily extracted from a given potential by studying the energy profile of two isolated atoms, a dimer, as a function of distance:
\begin{align}
    E_{2b,i}\left(\*r_i,\*r_j\right) &= V^{(2)}\left(\*r_i-\*r_j\right).
    \label{eq:e2b}
\end{align}
It is thus natural to ask whether this function might converge to the real dimer dissociation curve, or to another arbitrary function for a dataset that does not contain dimers.

While nonlinearities can improve the performance of machine learning potentials by representing higher-order terms when the expansion is truncated, care should be taken in how they are applied as certain applications may invalidate the body-ordering of the descriptor \cite{batatiaDesignSpaceE3equivariant2025}. In the present work, we do not aim to build the most accurate potentials, but to have a more grounded understanding of their behavior in the ACE framework. As such, we limit our investigations to training simpler linear models as well as not employing very high of a body-order where returns are diminished. Linear ACE models perform comparably well relative to neural network potentials, and have been successfully employed to describe molecular \cite{kovacsLinearAtomicCluster2021} and condensed phases \cite{lysogorskiyPerformantImplementationAtomic2021}.

\section{Methods}
\subsection{Datasets}\label{sec:datasets}
\begin{figure}
    \centering
    \includegraphics[width=0.5\linewidth]{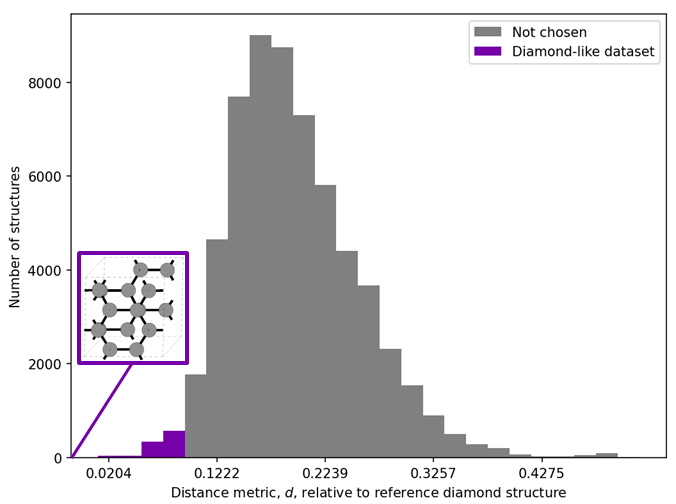}
    \caption{Histogram of fingerprint distances of the configurations in the Carbon dataset in \cite{shaiduSystematicApproachGenerating2021} relative to a reference diamond-like structure (inset). $1000$ configurations were chosen within a narrow region in structure space.}
    \label{fig:diahist}
\end{figure}

The datasets we use in the investigations consists of elemental structures of Carbon and Silicon. For Carbon, three datasets were used: the first set consists of $247$ dimer structures at distances between $0.5$~\AA\ and $8$~\AA, with half selected for the validation set. The other two are sets of $1000$ structures coming from a collection of Carbon allotropes at low pressures ($0-0.1$ GPa) \cite{shaiduSystematicApproachGenerating2021}. One subset contains a homogeneous collection of diamond-like structures clustered using a Euclidean distance measure \cite{oganovHowQuantifyEnergy2009}. The diamond-like dataset is characterized through a histogram of distances to a reference diamond structure in Fig.~\ref{fig:diahist}. The last subset are structures obtained using farthest point sampling \cite{eldarFarthestPointStrategy1997} to maximize its diversity. For the diamond-like and diverse datasets, a validation set was constructed from $1000$ diverse structures that have not been seen in training. The Carbon data were calculated using the \verb+Quantum Espresso+ package \cite{giannozziQUANTUMESPRESSOModular2009} using the rVV10 functional \cite{sabatiniNonlocalVanWaals2013} to account for nonlocal interactions.

Another dataset of molecular Silicon structures taken from \cite{chongPredictionRigiditiesDatadriven2025} was considered as simpler cases where the maximum body order $K$ in the bases is comparable to the number of neighbors in the structures. For each subset, $90\%$ was used for training and the rest were used for validation.

\subsection{Descriptor}\label{sec:descriptor}
We use a linear ACE descriptor as implemented by both the \verb+pacemaker+ package \cite{bochkarevEfficientParametrizationAtomic2022} and the \verb+ACEpotentials.jl+ \cite{wittACEpotentialsjlJuliaImplementation2023a}. We start from $K=2$ body functions and progressively add more terms of the same order. We validate the models at each addition of basis functions, and monitor when the interim model begins to overfit. Before the generalization error increases, we then introduce higher order $K=3$ terms while keeping the number of lower order terms fixed, and then repeat the procedure for higher orders. ACE descriptors employ spherical harmonics to describe angular functions, with $n,l$ values chosen to respect rotational symmetry. Low $n,l$ values were chosen to avoid overfitting the model from too many basis functions.

\subsection{Fitting procedure} \label{sec:initial_tests}
Training was done with a least squares error method for both \verb+pacemaker+ and \verb+ACEpotentials.jl+ methods. For \verb+pacemaker+ the loss $\mathcal{L}_p$ is,
\begin{align}
    \mathcal{L}_p =\left(1-\kappa \right)\Delta^2_E + \kappa\Delta^2_F + \Delta_{\text{coeff}} + \Delta_{\text{rad}}
    \label{eq:loss}
\end{align}
where $\Delta^2_E$ and $\Delta^2_F$ are the corresponding energy and force root-mean-square errors (RMSE), and $\kappa$ is the coefficient that balances the energy and force contributions to $\mathcal{L}$. The term $\Delta_{\text{coeff}}$ regularizes the basis function coefficients\footnote{This term includes the L1 and L2 regularizations.}, while $\Delta_{\text{rad}}$ regularizes the coefficients of the radial components of the basis function. \verb+ACEpotentials.jl+ minimizes the loss $\mathcal{L}_A$ through the kernel ridge regression methods,
\begin{equation}
    \mathcal{L}_A = ||\mathbf{W}\left(\mathbf{y}-\mathbf{Ac}\right)||^2 + \lambda||\mathbf{\Gamma c}||^2
    \label{eq:loss_acejl}
\end{equation}
where $\mathbf{W}$ encodes the relative loss weightings for energies and forces, $\mathbf{y}$ contains the labeled data, $\mathbf{A}$ is the basis matrix and $\mathbf{c}$ are the parameters to be learned to minimize $\mathcal{L}_A$. $\mathbf{\Gamma}$ contains a more general treatment to regularization allowing different penalizing strengths to the coefficients $c$, and $\lambda$ dictates the strength of the regularization. We used the Bayesian Linear Regression to let the model dictate appropriate values for $\lambda$ and $\mathbf{\Gamma}$. Details of the regression methods are found in \citet{wittACEpotentialsjlJuliaImplementation2023a}, and the Supporting Information discusses more details in the numerical setup. Evaluation of dimer curves were done using the Atomic Simulation Environment (ASE) by \citet{hjorthlarsenAtomicSimulationEnvironment2017}.

Aside from the parameters involving basis set complexities, radial bases/geometric priors, and the loss weightings, other values were set to default for both \verb|pacemkaer| and \verb|ACEpotentials.jl| to ensure a reasonable set of controlled parameters. 
\section{Results and Discussion}
\subsection{Self-interacting ACE bases}
\begin{figure}
    \centering
    \includegraphics[width=0.9\linewidth]{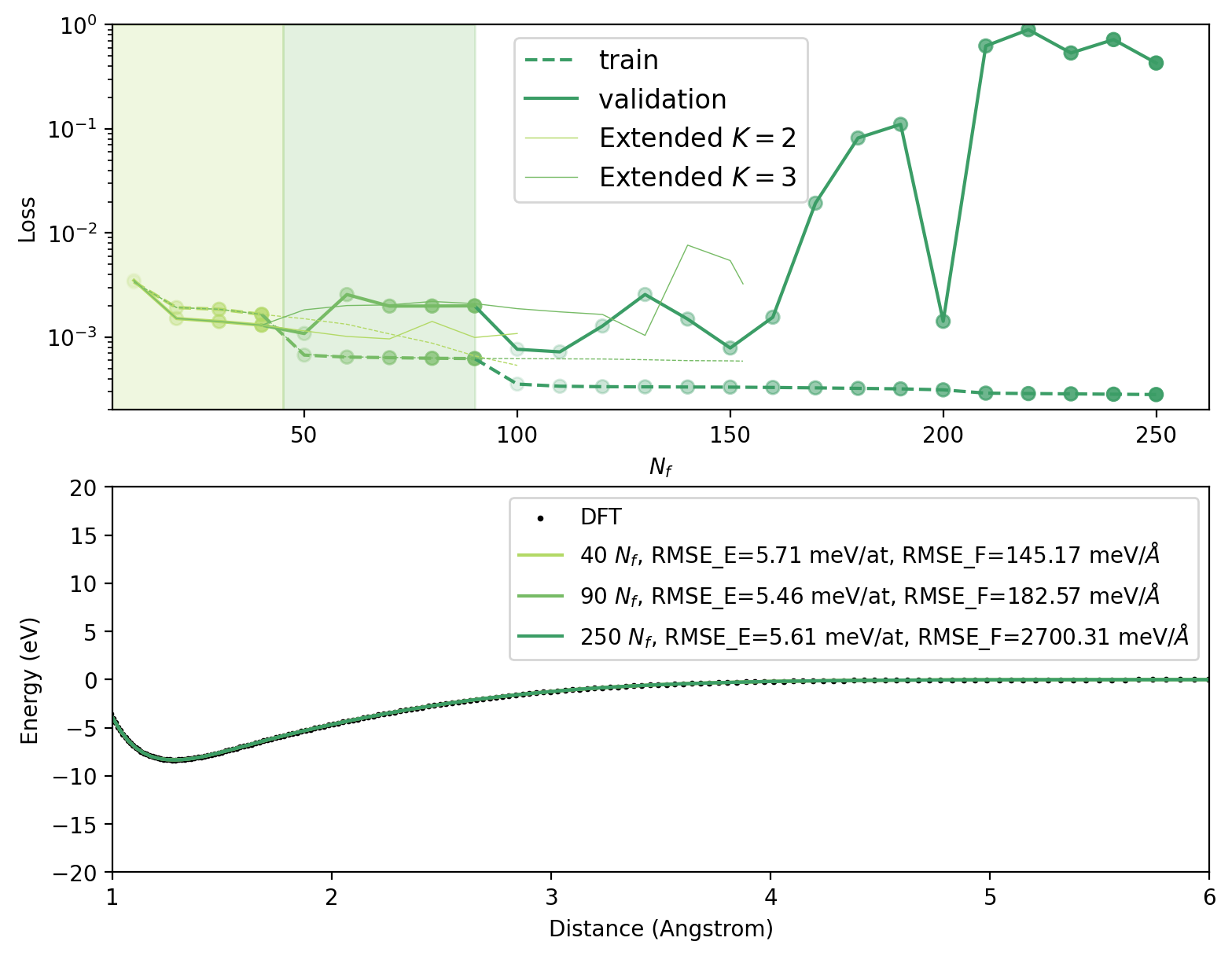}
    \cprotect\caption{\verb|pacemaker| (Top) Loss as a function of number of basis functions, $N_f$, for the dimer dataset. Training and validation errors as dashed and solid curves. Curve colors indicate potentials built from basis sets which are exclusively two-body (light green), up to three-body (green), and up to four-body functions (dark green), respectively. Light solid lines indicate the loss if more functions were introduced without introducing higher $K$. (Bottom) dimer curves for each of the potentials trained at certain $N_f$, with reference DFT data as points. Heavy lines represent the potential with the largest $N_f$ before functions of higher $K$ are introduced.}
    \label{fig:results_ogdim}
\end{figure}
We start by fitting the dimer dataset with an increasing number of basis functions, $N_f$, up to $250$ elements and $K=4$ body-order. The loss and dimer curves are shown in Fig.~\ref{fig:results_ogdim}, where light green, green, and dark green indicate data from potentials trained using $K$ up to 2, 3, and 4, respectively. The top plot shows the dependence of the loss function $\mathcal{L}$ on $N_f$. The validation loss (solid curve) converges for $K=2$ and increases with the addition of higher order functions, jumping two orders of magnitude as more $K=4$ functions are added. This is of course to be expected, as the dataset consists solely of 2 body structures. The bottom plot highlights the dimer curves of the potentials trained from the dimer set at the highest $N_f$ of a particular order $K$. We see that all potentials are able to reproduce the dimer curve as expected from the DFT results. 

The energy RMSE for the representative potentials are around $5$ meV/at, and force RMSE are slightly less than $200$ meV/\r{A} for $K=2$ and $K=3$. The force RMSEs increase to $2700$ meV/\r{A} for the overfitting potential at $N_f=250$, and are due to the short-range region of the curve having energies up to almost two orders of magnitude higher, making the deviations from DFT results more sensitive to errors. A comparison of the forces at the short range region may be found in the Supporting Information B.1.

\begin{figure}
    \centering
    \includegraphics[width=0.9\linewidth]{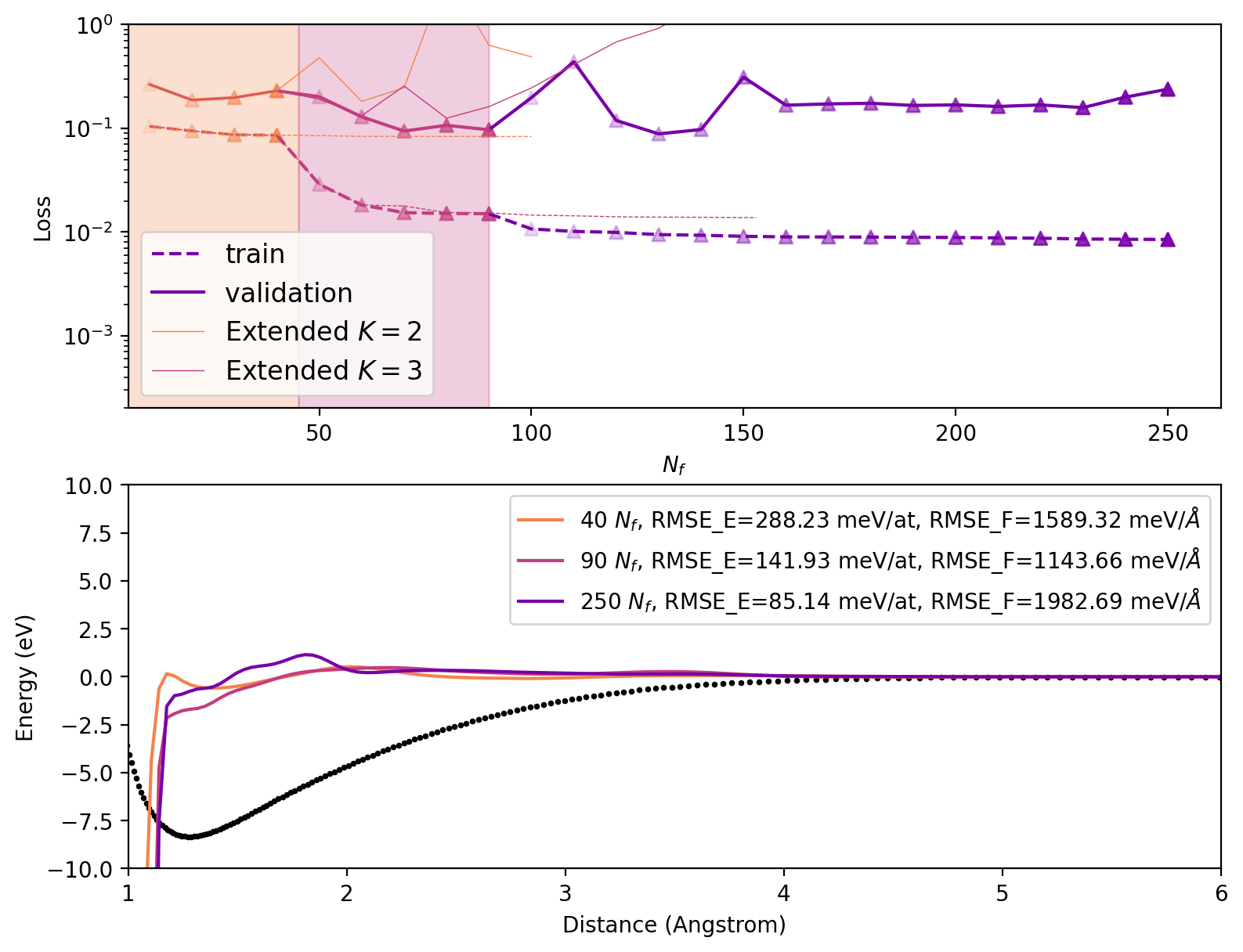}
    \cprotect\caption{\verb|pacemaker| (Top) $\mathcal{L}$ vs. $N_f$ and (Bottom) dimer curves for the homogeneous, diamond-like dataset. Orange and pink plots correspond to potentials with only $K=2$ and up to $K=3$ functions, respectively while purple curves include $K=4$ functions. The x-values of the dimer curve start at the minimum distance found in the homogeneous dataset.}
    \label{fig:results_ogdia}
\end{figure}

The results for the homogeneous diamond-like dataset are shown in Fig.~\ref{fig:results_ogdia}. Unlike the potentials trained from the dimer dataset, the training loss from the diamond-trained potentials reduce by an order of magnitude upon the introduction of the $K=3$ functions. The light orange line indicates that given the same $N_f$, the basis set purely constructed from $K=2$ functions does not have as pronounced of a decrease in $\mathcal{L}$. We see that the validation over the diverse set has losses improving for $K=3$, but overfitting occurs at $K=4$. The dimer curves from these potentials show a considerable amount of variability even after the loss is converged, and they remain far from the expected dimer behavior dictated by DFT. Of course, the lack of diversity in this dataset, leading to very small variations in the radial distribution function, might be part of the reason why a rather flat 2-body curve is obtained. Moreover, the overfitting for the large descriptor size could contribute to the variability of this contribution.

\begin{figure}
    \centering
    \includegraphics[width=0.9\linewidth]{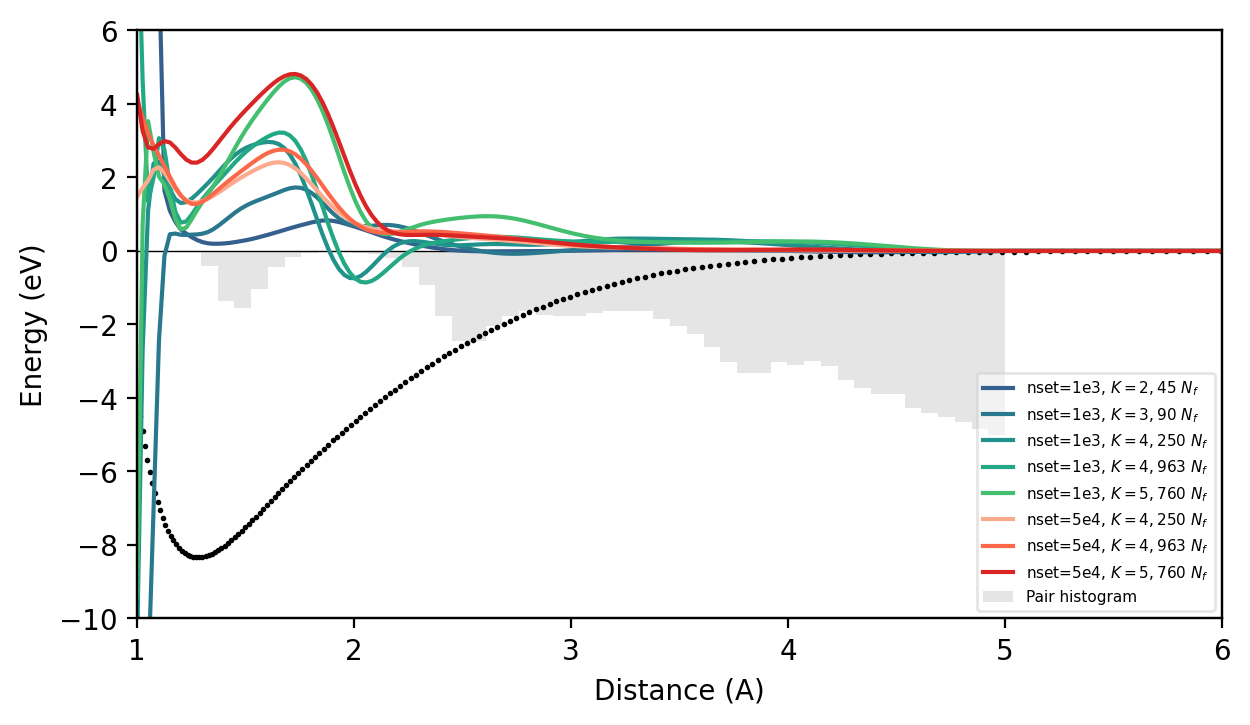}
    \cprotect\caption{\verb|pacemaker| Dimer curves for the diverse $1000$-structure (blue-green) and $50 000$-structure (orange-red) for various descriptor sizes. No convergence observed for both sets as higher $K$ functions are introduced.}
    \label{fig:results_dcurves_1k50k}
\end{figure}

We then look at the dimer curves of the potentials for different descriptor complexities, ranging from $K=2$ to $5$. The blue-green curves are potentials trained with $1000$-structures while the orange-red curves are trained with a larger $50000$-structure set. Fig.~\ref{fig:results_dcurves_1k50k} shows these curves having higher variability compared to the ones produced by the diamond-like dataset. While the curves look similar as more functions \textit{of the same $K$} are introduced in the set, as seen in the $K=4$ potentials, the curves of different $K$ limits do not seem to converge even when five-body contributions are already added. The variability is much more pronounced, even among potentials with similar maximum $K$ and more so when adding new terms. 

We note that the lack of pairs around the $2.0$ \r{A} region are due to the structures in the dataset belonging to relatively low pressures. The comparisons between the $1000$-structure and the more complete $50000$-structure datasets have virtually similar pair distribution functions due to the farthest point sampling methods done to sample the smaller subset. The pair distribution comparisons may also be seen in Supporting Information C. 

\begin{figure}
    \centering
    \includegraphics[width=0.9\linewidth]{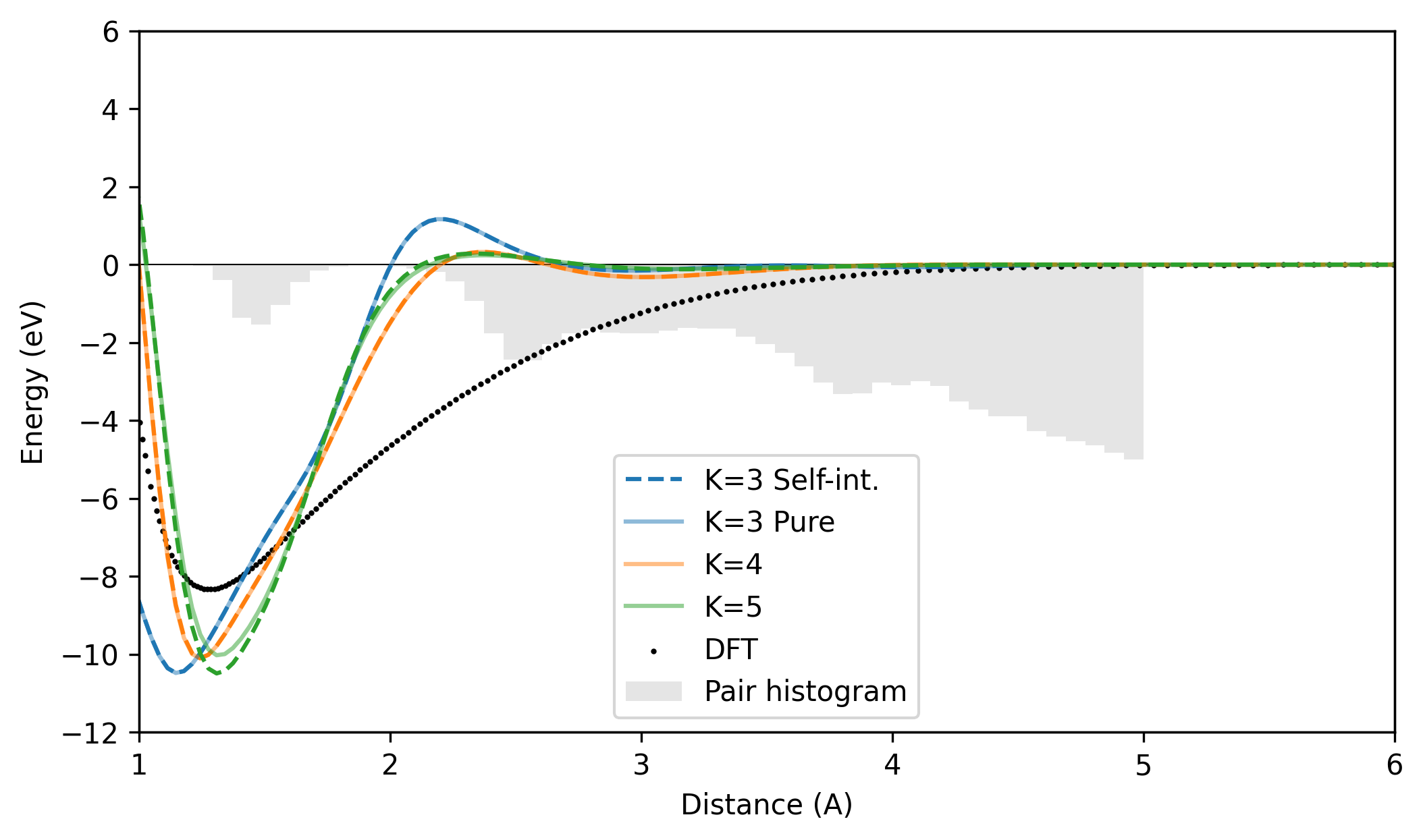}
    \cprotect\caption{\verb|ACEpotentials.jl| Dimer curves for $1000$-structure Carbon dataset trained with the Julia code \cite{wittACEpotentialsjlJuliaImplementation2023a}. Dashed lines indicate the ACE basis ($K=3,4,5$) with self-interactions similar to the one in pacemaker, while the solid lines are trained with a basis purified of self-interactions. Gray bars show the pair distance distribution in the training set. Using a purified basis for the Carbon dataset yields practically identical dimer curves with those from the self-interacting basis for up to $K=4$, with a qualtitative difference in the energy well for the $K=5$ potential.}
    \label{fig:results_acejl_ood_carbon}
\end{figure}

Finally, we try the same experiment using \verb|ACEpotentials.jl| using the $1000$-structure set with the plots. Fig.~\ref{fig:results_acejl_ood_carbon} shows qualitative improvements of the dimer curve in the equilibrium bond length $r_0$ and energy minimum may be attributed to \verb|ACEpotentials.jl| requiring additional user inputs (i.e. geometric priors) and more rigorous regularization schemes\cite{wittACEpotentialsjlJuliaImplementation2023a}. The routines involving Bayesian Linear Regression also allow for the automatic selection of regularization strengths $\lambda$ in Eq.~\ref{eq:loss_acejl} which helps reduce the number of hyperparameter sweeps done. In contrast, the absence of a more involved radial basis construction\footnote{The radial basis construction of pacemaker focuses on inner cutoffs that penalize oscillatory behavior at the short-range regions where data may be lacking, but optimizing this region of the curve is outside the study scope.} in the qualitative features of the dimer curve lead to its greater reliance on a richer dataset. 

Still, it seems that the potentials made from purified bases (solid curves) do not produce a marked improvement over the self-interacting basis (dashed lines) as they are mostly overlapping. While it seems that no improvements are seen in the purified basis, the study of the gram matrix conditioning from \cite{hoACEwoselfinteraction2024} indicates that the potentials become less stable when the total number of relevant particles $J$ (or neighbors in periodic systems) far exceed the maximum body-order, $K$, of the basis used to describe the potential\footnote{The notation of \citeauthor{hoACEwoselfinteraction2024} uses $N$ for the body-order variable instead of $K$, and uses $J$ to indicate the number of particles or neighbors considered in the system.}. 

\begin{table}[t]
\caption{Comparison of linear ACE models to neural network potentials in MAE for energies and forces for the Carbon dataset in \cite{shaiduSystematicApproachGenerating2021} for datasets containing $n_\text{set}=1000$ and $50000$ structures. Energies (E) are in meV/atom and force components (F) are in meV/\r{A}.}
\begin{tabular}{ll|l||l|l}
                                    & \multicolumn{2}{l||}{$n_\text{set}=1000$} & \multicolumn{2}{l}{$n_\text{set}=50000$} \\ \hline
\multicolumn{1}{l|}{Linear Potentials}     & E                   & F              & E                   & F               \\ \hline \hline
\multicolumn{1}{l}{pacemaker} & \multicolumn{4}{l}{} \\ \hline
\multicolumn{1}{l|}{$K=4, N_f=250$} & 48.50               & 241.27         & 44.06               & 222.42          \\
\multicolumn{1}{l|}{$K=4, N_f=493$} & 18.30               & 199.23         & 17.55               & 176.48          \\
\multicolumn{1}{l|}{$K=4, N_f=930$} & 15.40               & 167.73         & 15.27               & 166.70          \\
\multicolumn{1}{l|}{$K=5, N_f=793$} & 14.88               & 160.78         & 14.74               & 159.80          \\ \hline
\multicolumn{1}{l}{ACEpotentials.jl} & \multicolumn{4}{l}{} \\ \hline
\multicolumn{1}{l|}{$K=3, N_f=297$} & 38.10 & 336.84 & - & - \\
\multicolumn{1}{l|}{$K=4, N_f=802$} & 18.00 & 210.14 & - & - \\
\multicolumn{1}{l|}{$K=5, N_f=929$} & 16.72 & 202.78 & - & - \\ \hline \hline
\multicolumn{1}{l}{Neural Networks} & \multicolumn{4}{l}{} \\ \hline
\multicolumn{1}{l|}{PANNA small}    & 22.40               & 248.85         & 17.76               & 162.37          \\
\multicolumn{1}{l|}{PANNA mid}      & 20.93               & 241.36         & 12.41               & 137.45          \\
\multicolumn{1}{l|}{PANNA big}      & 17.88               & 242.02         & 8.21                & 166.95          \\
\multicolumn{1}{l|}{NequIP $l=1$}   & 20.40               & 213.00         & 7.96                & 110.00          \\
\multicolumn{1}{l|}{NequIP $l=2$}   & 6.83                & 105.00         & 5.21                & 51.40           \\
\multicolumn{1}{l|}{MACE}           & 6.81                & 103.14         & 1.82                & 51.31           \\ \hline \hline
\end{tabular}
\label{tab:results_PANNA2}
\end{table}

We also make sure that the performance of our linear models are reasonable by comparing their loss metrics to the performance of different neural network potentials trained on the same dataset in \citet{pellegriniPANNA20Efficient2023}, as shown in Table \ref{tab:results_PANNA2}. For the $1000$-size subsets, we found that the linear ACE models with the most complex descriptor sets ($K=4, N_f=930$ and $K=5, N_f=793$) were able to outperform all PANNA potentials and even the $l=1$ version of the NequIP \cite{batznerE3equivariantGraphNeural2022} potential. While promising, we may attribute the poor performance of the neural network potentials from the fact that $1000$ structures may not be enough data and the resulting potentials may overfit without appropriate regularizations.

However, they perform relatively worse with all other potentials when the complete $50000$-size Carbon set is used as the training set, which suggests that the nonlinearities absent in the ACE models provide a considerable improvement in training more complex datasets. Despite this, we believe that the inability of the linear models to capture the effective two-body interaction and converge to a defined shape are not due to the poor $\mathcal{L}$, but rather the nature and truncation of the expansion employed in the descriptor sets, which is further discussed in the experiments with purified bases for the Silicon dataset.

\subsection{Purified ACE bases}
\begin{figure}
    \centering
    \includegraphics[width=\linewidth]{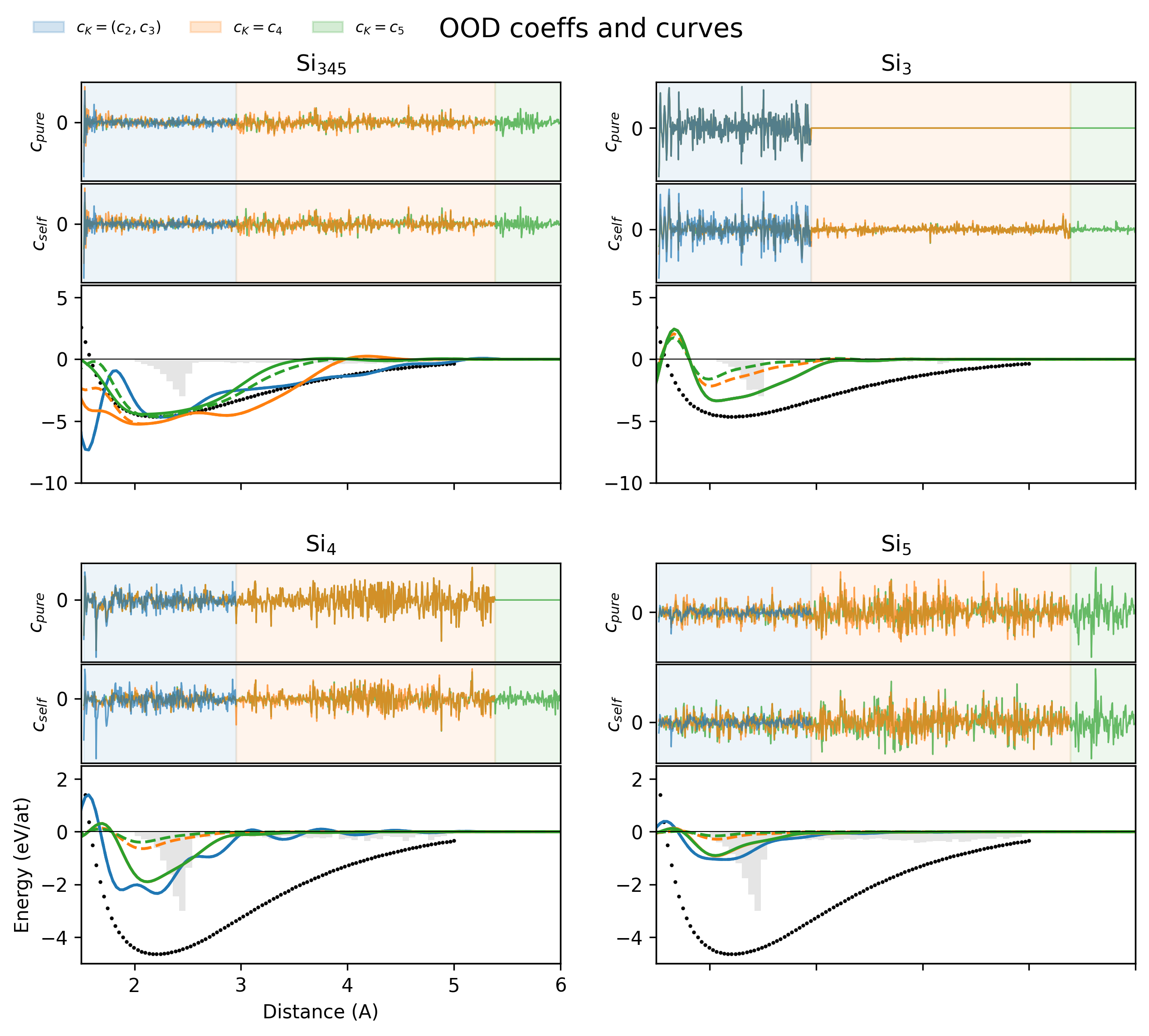}
    \cprotect\caption{\verb|ACEpotentials.jl| (Bottom of each quadrant) Dimer curves of potentials from the Silicon structures in \citet{chongPredictionRigiditiesDatadriven2025} trained at different subsets. Si$_{mn}$ indicate potentials were trained from datasets that include $m$-mers and $n$ mers, black points show the DFT dimer energies, and gray bars indicate the distributions of pairs in the dataset. Overlapping solid orange ($K=4$) and green ($K=5$) lines indicate that potentials with purified bases eventually converge to a defined shape when the total number of particles $N$ match the maximum body order $K$. (Upper plots of each quadrant) Coefficients of the purified ($c_{pure}$) and self-interacting ($c_{self}$) potenials. Blue, orange, and green lines correspond to coefficients of potentials capped at $K=3$, $K=4$, and $K=5$ body-ordered functions while blue, orange, and green shaded regions indicate coefficients corresponding to $K=2,3$, $K=4$, and $K=5$ body-ordered functions. Overlapping lines for a certain region mean that potentials have very similar basis coefficients, suggesting the expansion has stabilized for that particular body-order.}
    \label{fig:results_acejl_ood_silicon}
\end{figure}

We then repeat the same investigation on a dataset of Silicon structures\cite{chongPredictionRigiditiesDatadriven2025} containing dimers, trimers, tetramers, and pentamers so the maximum number of atomic neighbors matches $K=5$. We train potentials from subsets of $100$ trimers (Si$_3$), $100$ tetramers (Si$_4$), and $500$ pentamers (Si$_5$), and a mixture of the structures leaving out the dimers (Si$_{345}$) to mimic the diverse Carbon set. Fig.~\ref{fig:results_acejl_ood_silicon} highlights that the out-of-distribution Silicon potentials, those trained without dimers, also have curves that are not qualitatively recovered. Additionally, the Si$_{345}$ potentials, which strikes the closest parallel to the diverse Carbon dataset, exhibit the overlapping of purified and self-interacting potentials albeit at a lesser degree than the Carbon tests. When we look at the coefficients of the expansion, the non-overlap of the lines show the instability of the solutions for the potentials at different $K$. 

When focusing on the purely trimer trained Si$_3$ potentials, we see signatures of convergence at the purified basis. Dimer curves of potentials at all $K$ overlapping and the basis coefficients $c_{pure}$ for the $2$- and $3$- body functions (in the blue region) are consistent with each other. The $c_{pure}$ corresponding to $K\geq 4$ functions are $0$, which suggest that the self-interactions are indeed removed. In the Si$_4$-trained potentials, dimer curves for the $K=3$ basis set already deviates from the others due to the lower-order $c_{pure}$ misrepresenting the higher-order tetramer interactions. We also see this happening for the Si$_5$-trained potentials and even for the $K=4$ basis sets for their misrepresentation of the $5$-body interactions. Overall, the qualitative features of the dimer curves do not seem to converge as basis functions become more complex with higher $K$. This makes it difficult to apply the potentials to structures it has not seen during training.

\begin{figure}
    \centering
    \includegraphics[width=\linewidth]{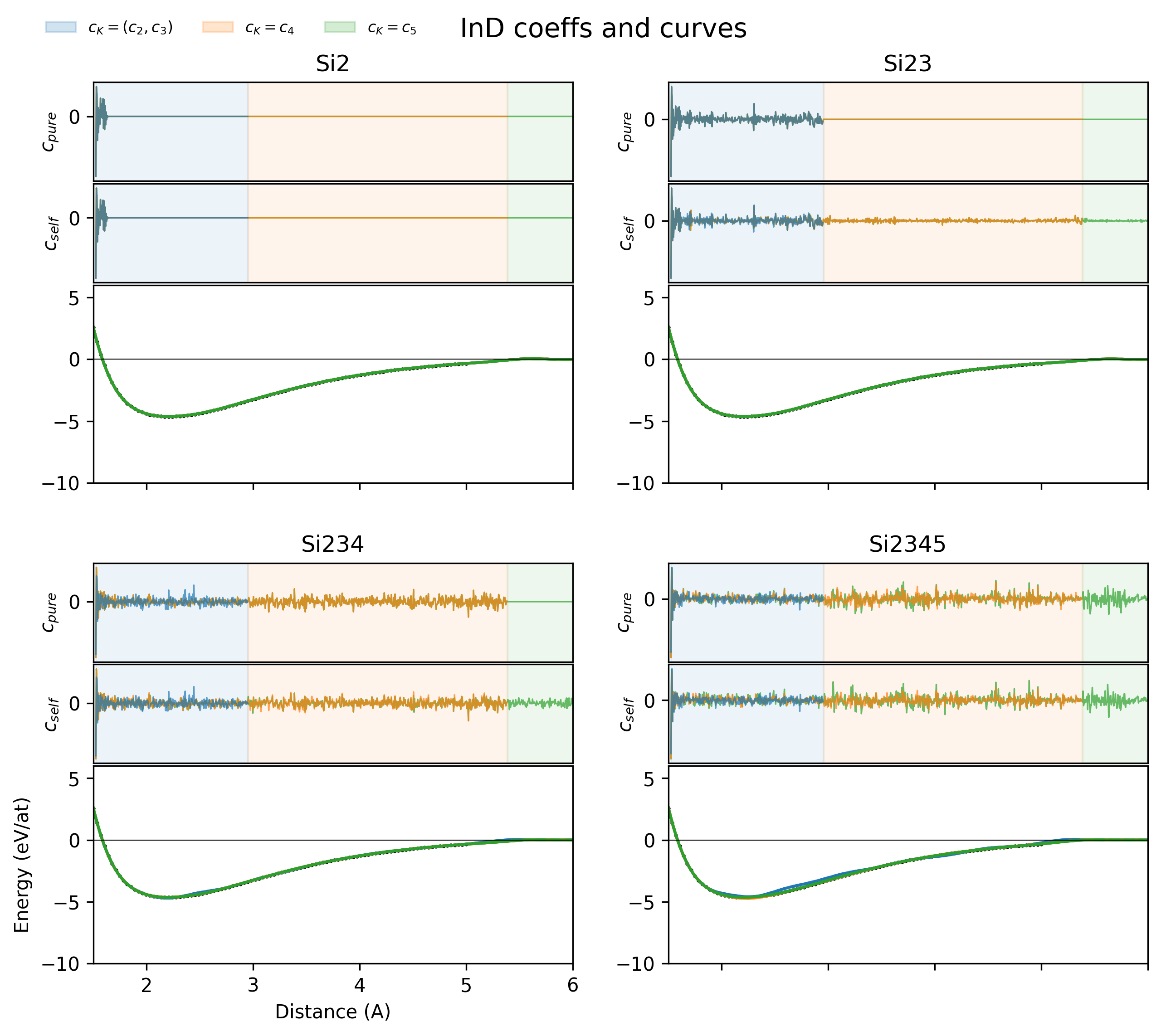}
    \cprotect\caption{\verb|ACEpotentials.jl| (Top of each quadrant) ACE coefficients and (bottom of each quadrant) dimer curves of Silicon potentials trained at different subsets, including dimers in all of them. All dimer curves recover the correct DFT values, and purified potentials. All $c_{self}$, except for the Si$_2$ training, had higher-order renormalizations as seen in the nonzero $K=3,4$ for Si$_{23}$ and $K=4$ for Si$_{234}$. $c_{pure}$ coefficients are similar for the $K=2,3$ terms in Si$_{23}$-trained potentials, and similar for $K=4$ terms in the Si$_{234}$ potentials.}
    \label{fig:results_acejl_ind_silicon}
\end{figure}

While generalization to the correct DFT curve seems unlikely, we can still see how the coefficients of purified linear ACE models perform when doing interpolative tasks and dimers are included in the training set. It is evident in Fig.~\ref{fig:results_acejl_ind_silicon} that the correct DFT results are recovered regardless of basis max $K$ or whether the basis was purified or not. To understand whether the solution we obtain is stable, we again look at the coefficients of the potentials and see that the purification is successful in disentangling lower-order terms from the contributions of functions at higher $K$. It is also worth noting that the $c_{self}$ for the Si$_2$-trained datasets look identical to the purified coefficients, and is possibly due to the fact that the self-interaction terms are not needed to capture the dimer energetics. 

\begin{figure}
    \centering
    \includegraphics[width=0.7\linewidth]{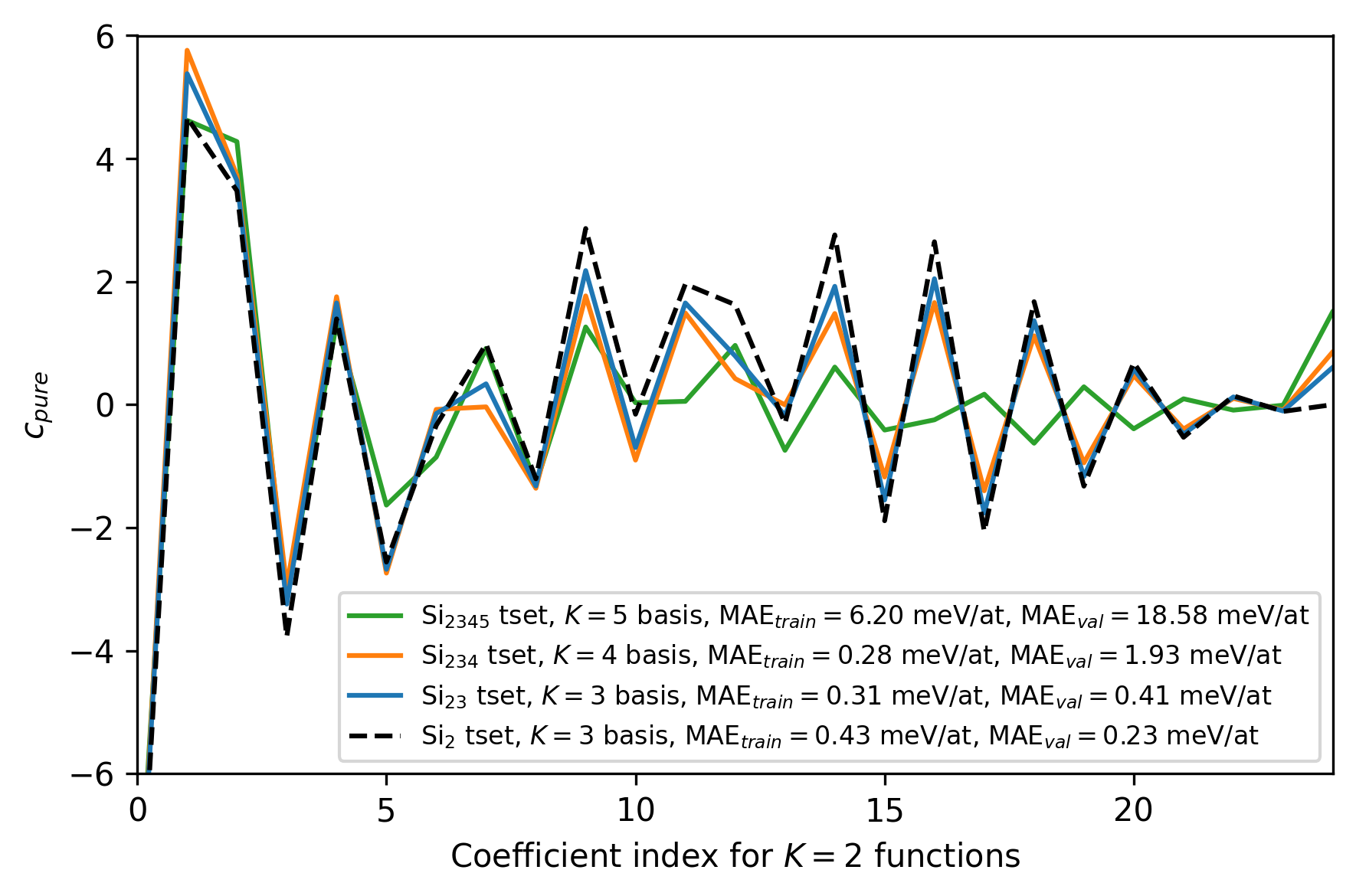}
    \cprotect\caption{\verb|ACEpotentials.jl| $c_{pure}$ of the $K=2$ basis functions for potentials with basis complexities matching the max number of particles, $K=J$.}
    \label{fig:results_acejl_ind_silicon_2bcoeffs}
\end{figure}

The mismatch in the overlap of $c_{pure}$ can still be seen in the $K=2,3$ functions of the Si$_{234}$ sets and the $K=2,3,4$ functions of the Si$_{2345}$ sets which is the signature of aliasing that occurs when the basis is not complex enough to capture the higher-order energetics. To see if we are consistently getting not only the correct dimer curve, but a stable pair potential, we look at the $c_{pure}$ of the $K=2$ basis functions for datasets where max $K$ matches the maximum number of particles $J$ in Fig \ref{fig:results_acejl_ind_silicon_2bcoeffs}. We include the Si$_2$ potential with the $K=3$ basis as the reference coefficient set. Surprisingly, we see that no other potential was able to capture the reference coefficients completely, with greater deviations coming from the Si$_{234}$ and Si$_{2345}$ potentials. Looking at the discrepancy between training and validation errors between potentials, the high validation error for the Si$_{234}$ potential suggests that more tetramers need to be sampled. For the Si$_{2345}$ potential, both training and validation errors being high implies that the $K=5$ basis functions as well as more pentamers are needed.

\section{Summary and conclusions} \label{sec:conclusions}
In this work, we studied the convergence of linear cluster expansions with respect to the body-ordered basis functions through the effective two-body interactions, $E_{2b}$ of elemental Carbon and Silicon systems. We used the atomic cluster expansion (ACE), with descriptor sets truncated at body-orders $K=2$ to $K=5$ trained on a diverse Carbon dataset and on Silicon dimers to pentamers. The potentials trained on diverse structures with standard ACE bases were not able to recover the correct dimer curves much less produce stable curves or solutions. While employing ACE bases removed of self-interactions still did not generalize to the DFT-expected results, properly tailored datasets and basis sets were able to show signs of convergence and stability in the curves and expansions. Stabilizing the coefficients of the expansion order-by-order by procedurally training potentials at bases up to $K$ from training sets with $K$-mers, and possibly utilizing uncertainty quantification metrics as detailed in \citet{chongPredictionRigiditiesDatadriven2025}, shows promise as a framework to build a potential with a more interpretable basis using the cluster expansion.

\section*{Supporting Information Available}
Detailed information on the simulation parameters for both codes is found in Sec. A. Hyperparameter investigations, particularly in balancing the loss contributions between energies and forces, are found in Sec. B. Additional plots may be found in Sec. C, namely force plots between DFT and dimer-trained potential, loss analysis of diverse dataset dimer curves, biasing the diverse dataset with dimer data, learning curves of the diverse dataset runs as well as the extended datasets. Datasets and codes used for calculations may be found in \href{https://github.com/apolmiguel/aceconverge2025}{https://github.com/apolmiguel/aceconverge2025}.

\section*{Acknowledgement}
AT thanks Gabriele Coiana and Jerry Ho for fruitful discussions. This study was funded by the European Union – NextGenerationEU – PNRR M4C2-I.1.4, in the framework of the project CN-HPC: National Centre for HPC, Big Data and Quantum Computing - Spoke 7  - (MUR ID: CN\_00000013 – CUP: G93C22000600001).
FP work was supported by EU – FSE REACT-EU and Italian MUR under contract 45-I-D2100-1 of the PON-RI 2014-2020 initiative. SdG work was supported in part by the European Commission through the MAX Centre of Excellence for supercomputing applications (grant numbers 10109337 and 824143). Computational resources were provided by CINECA.

\appendix
\renewcommand{\thefigure}{S\arabic{figure}}

\section{Simulation Parameters}
For the \verb|pacemaker| tests, the cutoff radius is set to $5$ Angstrom. The potential does not use the nonlinear version of the Finnis-Sinclair potential which is set on by default in the \lstinline{pacemaker} package. The radial basis is the Chebyshev expansion with the cosine cutoff function. The energy-force cost balancing is set that the RMSE for energies is weighted 50 times more than the forces, and further investigations is shown in Section B. The optimizer is set to BFGS with a maximum of $3000$ iterations and a batch size of $100$. Details about the input files may be found at the \href{https://pacemaker.readthedocs.io/en/stable/pacemaker/inputfile/}{documentation} of \lstinline{pacemaker}. Table ~\ref{tab:runparams} distinguishes the training parameters done for the dimer curves presented in the main paper.
\begin{table*}[t]
    \caption{Simulation Parameters for binding curve experiments in pacemaker}
    \label{tab:runparams}
    \centering
    \begin{tabular}{llrlll}
    \multicolumn{1}{l|}{Training set}                 & \multicolumn{1}{l|}{Descriptor label} & \multicolumn{1}{l|}{Train steps} & \multicolumn{1}{l|}{Validation set}      & \multicolumn{1}{l|}{nmax}     & lmax    \\ \hline
    \multicolumn{1}{l|}{\multirow{3}{*}{Dimer $124$}} & \multicolumn{1}{l|}{$K=2, N_f=45$}    & \multicolumn{1}{r|}{3000}        & \multicolumn{1}{l|}{val\_dimer $123$}    & \multicolumn{1}{l|}{100}      & 0       \\
    \multicolumn{1}{l|}{}                             & \multicolumn{1}{l|}{$K=3, N_f=90$}    & \multicolumn{1}{r|}{3000}        & \multicolumn{1}{l|}{val\_dimer $123$}    & \multicolumn{1}{l|}{45/8}     & 0/2     \\
    \multicolumn{1}{l|}{}                             & \multicolumn{1}{l|}{$K=4, N_f=250$}   & \multicolumn{1}{r|}{3000}        & \multicolumn{1}{l|}{val\_dimer $123$}    & \multicolumn{1}{l|}{45/5/4}   & 0/2/2   \\ \hline
    \multicolumn{1}{l|}{Diamond $1000$}               & \multicolumn{1}{l|}{$K=2, N_f=45$}    & \multicolumn{1}{r|}{3000}        & \multicolumn{1}{l|}{val\_diverse $123$}  & \multicolumn{1}{l|}{100}      & 0       \\
    \multicolumn{1}{l|}{}                             & \multicolumn{1}{l|}{$K=3, N_f=90$}    & \multicolumn{1}{r|}{3000}        & \multicolumn{1}{l|}{val\_diverse $123$}  & \multicolumn{1}{l|}{45/8}     & 0/2     \\
    \multicolumn{1}{l|}{}                             & \multicolumn{1}{l|}{$K=4, N_f=250$}   & \multicolumn{1}{r|}{3000}        & \multicolumn{1}{l|}{val\_diverse $123$}  & \multicolumn{1}{l|}{45/5/4}   & 0/2/2   \\ \hline
    \multicolumn{1}{l|}{Diverse $1000$}               & \multicolumn{1}{l|}{$K=2, N_f=45$}    & \multicolumn{1}{r|}{3000}        & \multicolumn{1}{l|}{val\_diverse $1000$} & \multicolumn{1}{l|}{100}      & 0       \\
    \multicolumn{1}{l|}{}                             & \multicolumn{1}{l|}{$K=3, N_f=90$}    & \multicolumn{1}{r|}{3000}        & \multicolumn{1}{l|}{val\_diverse $1000$} & \multicolumn{1}{l|}{45/8}     & 0/2     \\
    \multicolumn{1}{l|}{}                             & \multicolumn{1}{l|}{$K=4, N_f=250$}   & \multicolumn{1}{r|}{3000}        & \multicolumn{1}{l|}{val\_diverse $1000$} & \multicolumn{1}{l|}{45/5/4}   & 0/2/2   \\
    \multicolumn{1}{l|}{}                             & \multicolumn{1}{l|}{$K=4, N_f=493$}   & \multicolumn{1}{r|}{3000}        & \multicolumn{1}{l|}{val\_diverse $1000$} & \multicolumn{1}{l|}{45/8/4}   & 0/3/3   \\
    \multicolumn{1}{l|}{}                             & \multicolumn{1}{l|}{$K=4, N_f=960$}   & \multicolumn{1}{r|}{3000}        & \multicolumn{1}{l|}{val\_diverse $1000$} & \multicolumn{1}{l|}{45/9/5}   & 0/6/3   \\
    \multicolumn{1}{l|}{}                             & \multicolumn{1}{l|}{$K=4, N_f=793$}   & \multicolumn{1}{r|}{3000}        & \multicolumn{1}{l|}{val\_diverse $1000$} & \multicolumn{1}{l|}{45/8/4/3} & 0/3/3/2 \\ \hline
    \multicolumn{1}{l|}{Diverse $50000$}              & \multicolumn{1}{l|}{$K=4, N_f=250$}   & \multicolumn{1}{r|}{3000}        & \multicolumn{1}{l|}{val\_diverse $1000$} & \multicolumn{1}{l|}{45/5/4}   & 0/2/2   \\
    \multicolumn{1}{l|}{}                             & \multicolumn{1}{l|}{$K=4, N_f=493$}   & \multicolumn{1}{r|}{810}         & \multicolumn{1}{l|}{val\_diverse $1000$} & \multicolumn{1}{l|}{45/8/4}   & 0/3/3   \\
    \multicolumn{1}{l|}{}                             & \multicolumn{1}{l|}{$K=4, N_f=960$}   & \multicolumn{1}{r|}{418}         & \multicolumn{1}{l|}{val\_diverse $1000$} & \multicolumn{1}{l|}{45/9/5}   & 0/6/3   \\
    \multicolumn{1}{l|}{}                             & \multicolumn{1}{l|}{$K=4, N_f=793$}   & \multicolumn{1}{r|}{1151}        & \multicolumn{1}{l|}{val\_diverse $1000$} & \multicolumn{1}{l|}{45/8/4/3} & 0/3/3/2 \\ \hline
                                                      &                                       & \multicolumn{1}{l}{}             &                                          &                               &         \\
                                                      &                                       & \multicolumn{1}{l}{}             &                                          &                               &        
    \end{tabular}
\end{table*}

\begin{figure}
    \centering
    \includegraphics[width=\linewidth]{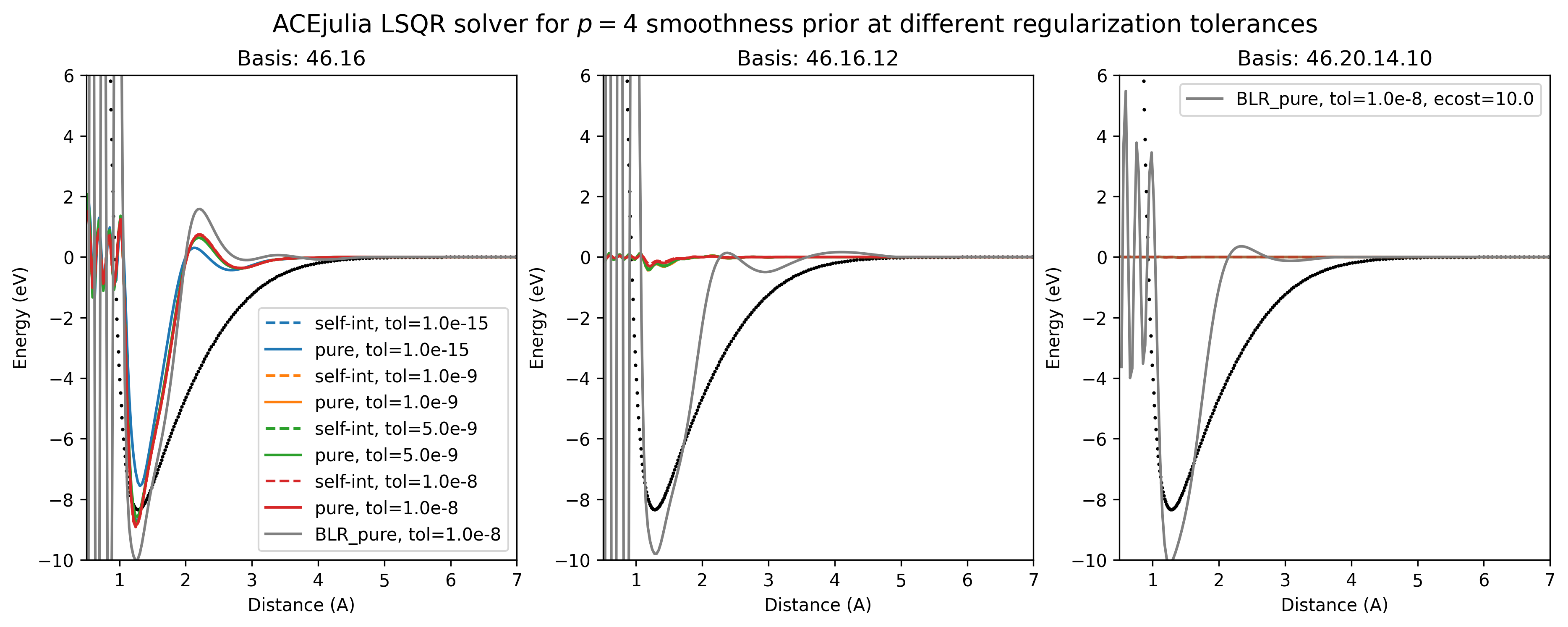}
    \caption{Comparison of ACEpotentials.jl dimer curves between LSQR (colored) and BLR (gray) solvers for basis sets $K=3$ to $K=5$ (left to right) and smoothness prior $p=4$. Different colors for potentials regressed with LSQR methods indicate different regularization tolerance strengths. $K=3$ limited LSQR-solved potentials reaches the BLR-solved potentials as the regularization tolerance is increased. $K=4,5$ LSQR-solved potentials had vanishing dimer curves for all tolerances. Energy-to-force loss weighting was set to $50$ for all potentials except for the BLR-solved $K=5$ which was set to $10$ due to issues in getting noninvertible matrices in the solver. }
    \label{fig:acejulia_lsqr}
\end{figure}

We used \verb|v0.6.7| of \verb|ACEpotentials.jl| as it was a stable version for the implementation of the purification algorithm. We used the \verb|ACE1x.ace_basis| module to create purified bases, and is done by setting \verb|pure = true| in the arguments. We chose to specifically control the maximum degrees per order, with the $K=3, 4, 5$ basis sets being $\left[24,20\right]$, $\left[24,20,16\right]$, and $\left[24,20,16,12\right]$ respectively. These led to each basis set having $297$, $802$, and $929$ basis functions, respectively\footnote{Each basis set had $24$ two-body, and $273$ three-body basis functions. $K=4,5$ had $505$ four-body functions while $K=5$ had $127$ five-body functions.}. The number of basis functions were not explicitly selected as the number of functions are consequences of the maximum orders selected by the \verb|ACE1x.ace_basis| function. We made sure to make the maximum degrees of each order even, with the lower orders having degrees considerably higher than the succeeding ones following the prescription of \citet{hoACEwoselfinteraction2024} to ensure that the purified bases span the same space as the original self-interacting basis. We chose \verb|rcut=5.0| and expected equilibrium bond length \verb|r0=1.287| for the Carbon tests as these are the expected parameters for the dataset, while \verb|rcut=5.0, r0=2.207| were chosen for the Silicon datasets. 

\verb|ACEpotentials.jl| training was done using the Bayesian Linear Regression module by using \verb|ACEfit.BLR()| to atuomatically determine the best values for the smoothness prior $\mathbf{\Gamma}$ and regularization strength $\lambda$. We also found out that this solver provided the most stable dimer curves with respect to changes in basis as compared to \verb|LSQR| solvers in \citet{wittACEpotentialsjlJuliaImplementation2023a} as seen in Fig.~\ref{fig:acejulia_lsqr}. Optimizations for LSQR-solved potentials was not done due to the vanishing dimer curves at higher $K$. BLR-solved potentials were used mostly ``out-of-the-box" values (i.e. Cholesky factorizations used), but qualitative similarities were already present for the \verb|ACEpotentials.jl| dimer curves. Careful hyperparameter sweeps can be done to optimize the losses, but this is not the main focus of the work. 

\section{Relative weighting for energy and force losses}
\begin{figure}
    \centering
    \includegraphics[width=\linewidth]{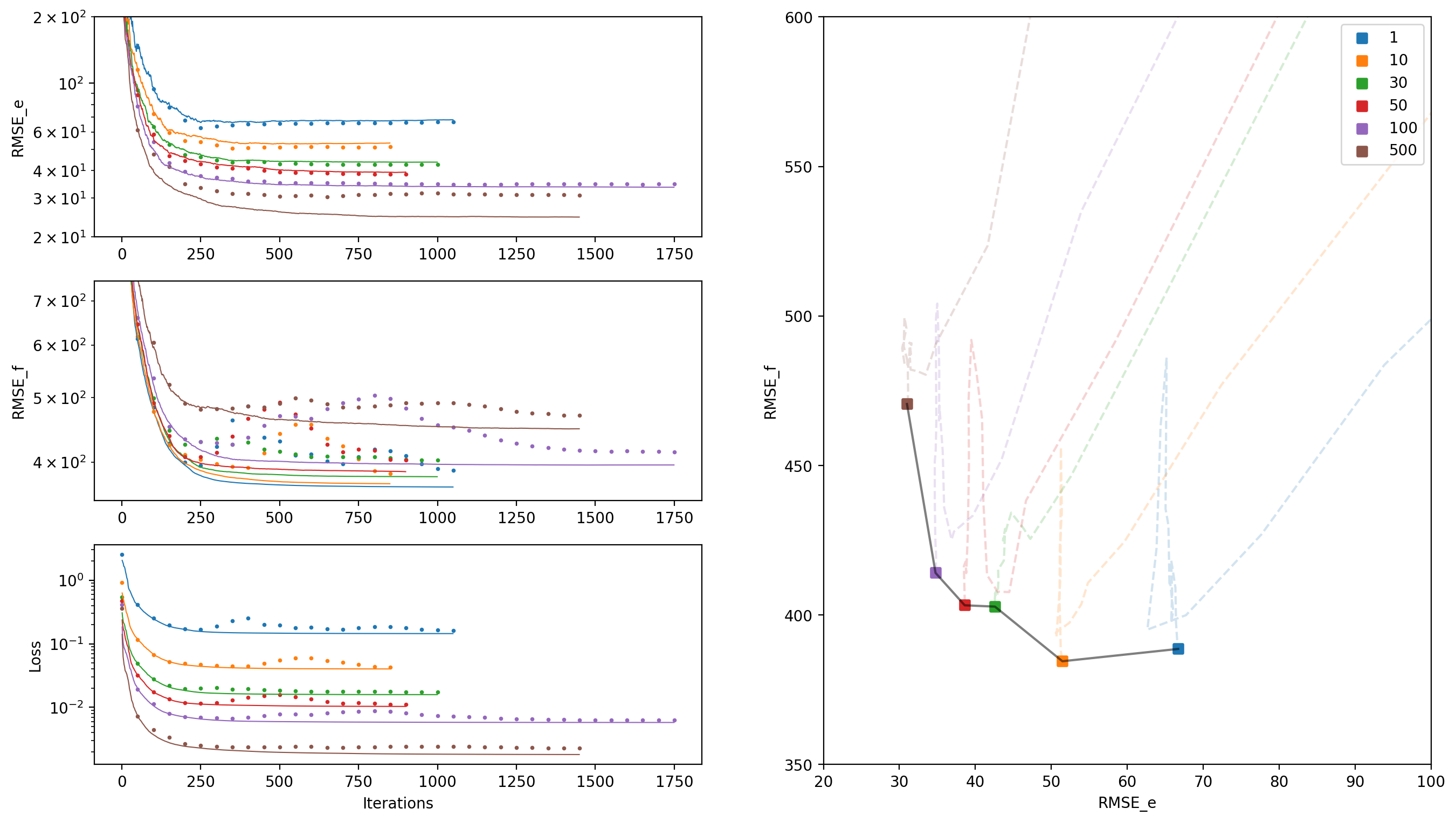}
    \caption{(Left, top to bottom) RMSE for energy, forces, and loss for different ratios for coefficients between energy and forces. $10$ means energy is weighted $10$ times more. (Right) RMSE forces vs. RMSE energies for the different loss coefficient ratios. Faint lines indicate the progression of each of the losses as a function of iterations. Final values are cut at the point where the validation error starts to increase considerably.}
    \label{fig:lossbalancing}
\end{figure}

In \verb|pacemaker|, we chose default values for the regularization weights but selected the appropriate energy-to-force loss weighting ratio with the one that simultaneously minimizes the RMSE for energies and forces. Visually, it is the point closest to the origin in the right plot of Fig.~\ref{fig:lossbalancing}. We see that the best results show for when the energy is weighted $50$ times stronger than forces in the regression task (red), and seems to be the start of the dimer curves being ``converged'' with respect to the coefficient ratio. The study focused on the energetics of the system, so weighting the energies seemed appropriate and our tests showed that increasing the coefficient from $1$ to $50$ had a marked improvement in the reduction of RMSE for energies, while only slightly penalizing the forces. 

\begin{figure}
    \centering
    \includegraphics[width=0.7\linewidth]{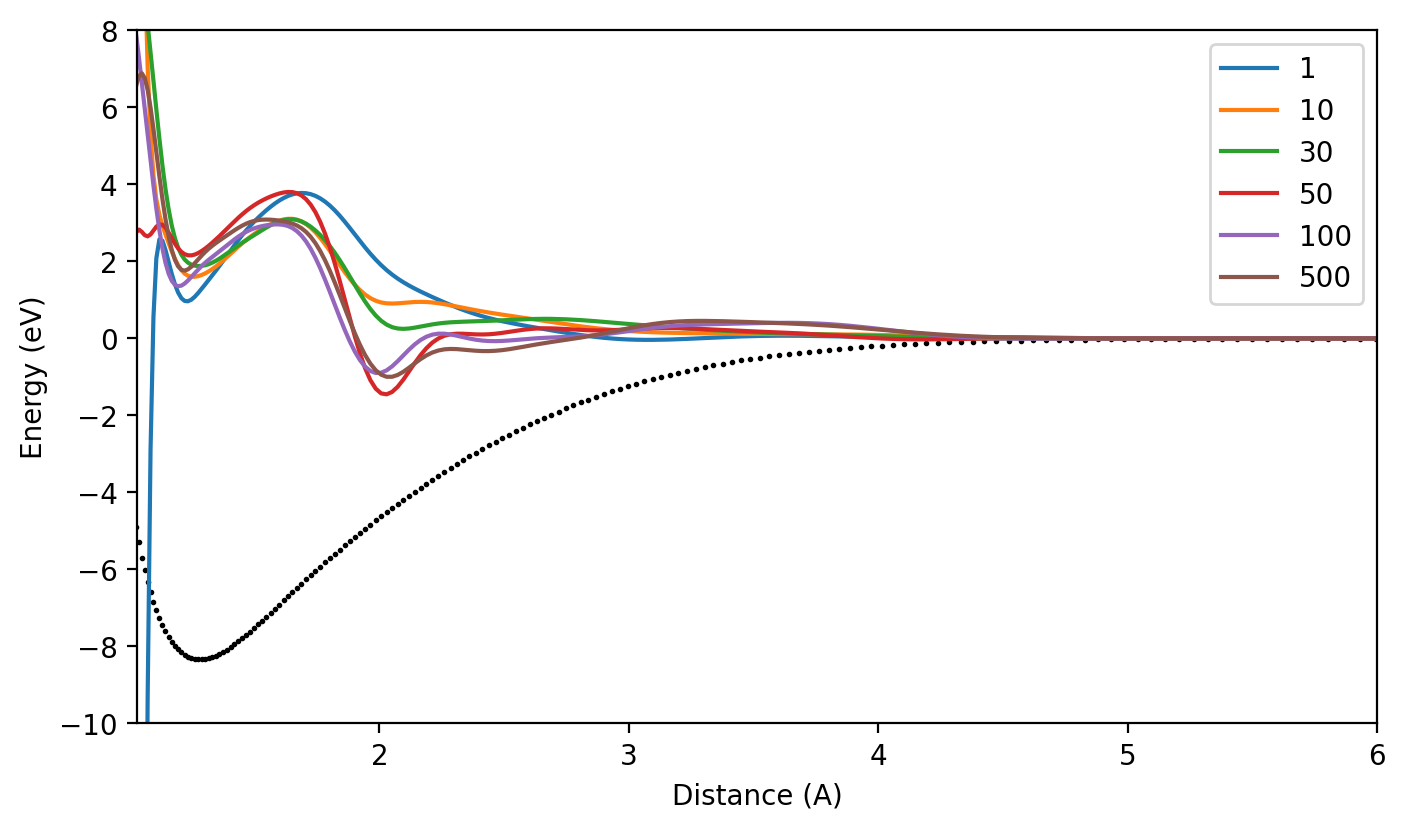}
    \caption{Dimer curves of each of the loss coefficient values at the points in the right of Fig.~\ref{fig:lossbalancing}. For the remainder of the tests in the main paper, the chosen energy loss coefficient is $50$ times the forces.}
    \label{fig:lossbalancing_dimers}
\end{figure}
We were not able to do the same loss balancing experiments for \verb|ACEpotentials.jl| due to problems in the matrix for the linear regression being non-invertible. We set the energy-force weighting for the Carbon dataset tests at $10$, while it was set at $50$ for the Silicon datasets. 

\newpage

\section{Additional plots}
\subsection{Force plots between DFT and dimer-trained potential}
\begin{figure}
    \centering
    \includegraphics[width=0.8\linewidth]{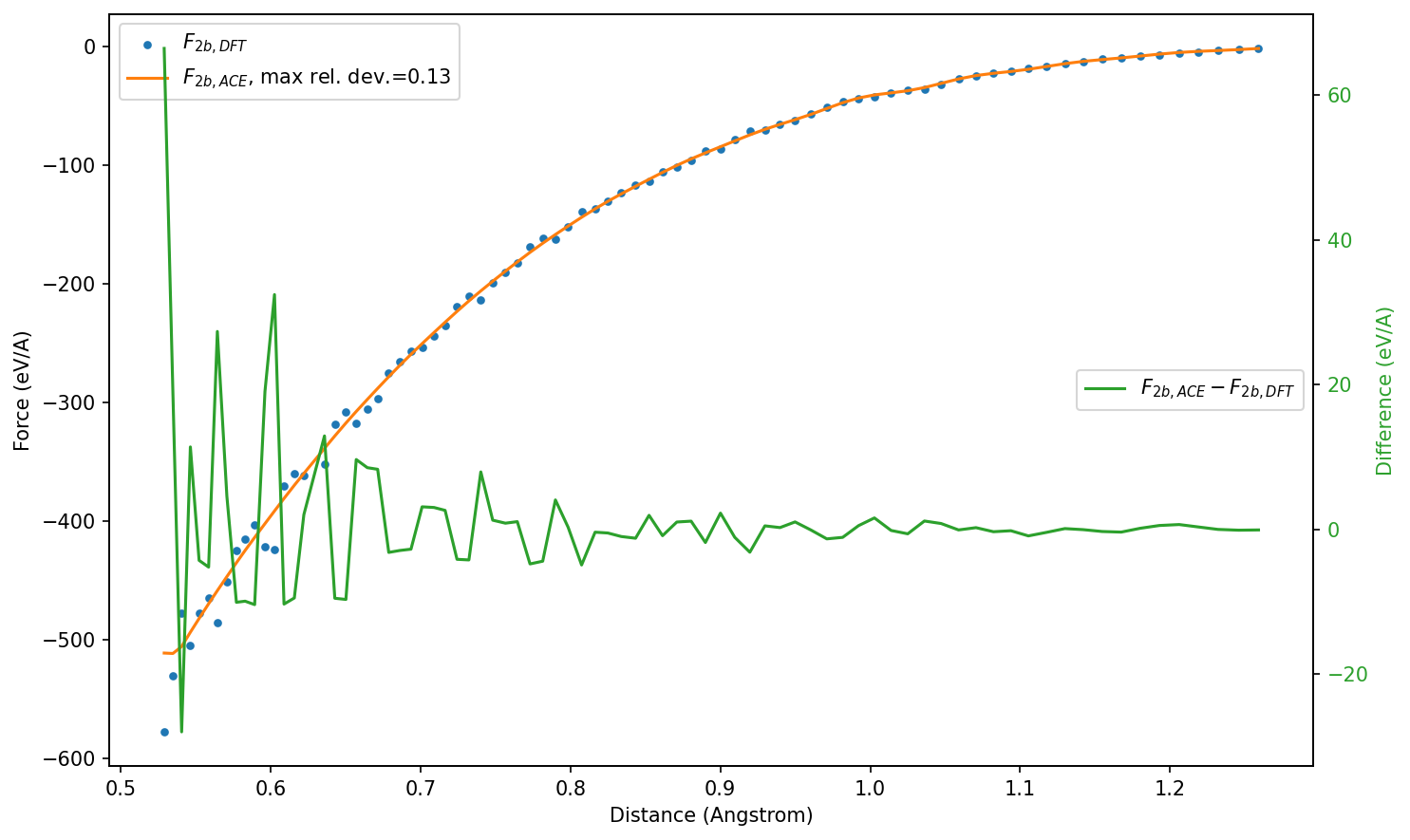}
    \caption{Forces at the short-range region for the dimer-trained potential and the difference between the forces of the DFT and the dimer-trained potential (green, right y-axis).}
    \label{fig:fdiff}
\end{figure}

Fig.~\ref{fig:fdiff} shows the force values from the reference DFT data (blue scatter) as compared with the values predicted by the NNP (orange line) at the short-range region for the dimer-trained potential with $K=4$ and $N_f=250$. We see that despite relatively low relative errors (maximum at $13\%$), the absolute value of the forces admit large errors which contribute to the large validation errors at the overfitting region. 
\newpage
\subsection{Loss analysis of diverse dataset dimer curves}
\begin{figure}
    \centering
    \includegraphics[width=0.7\linewidth]{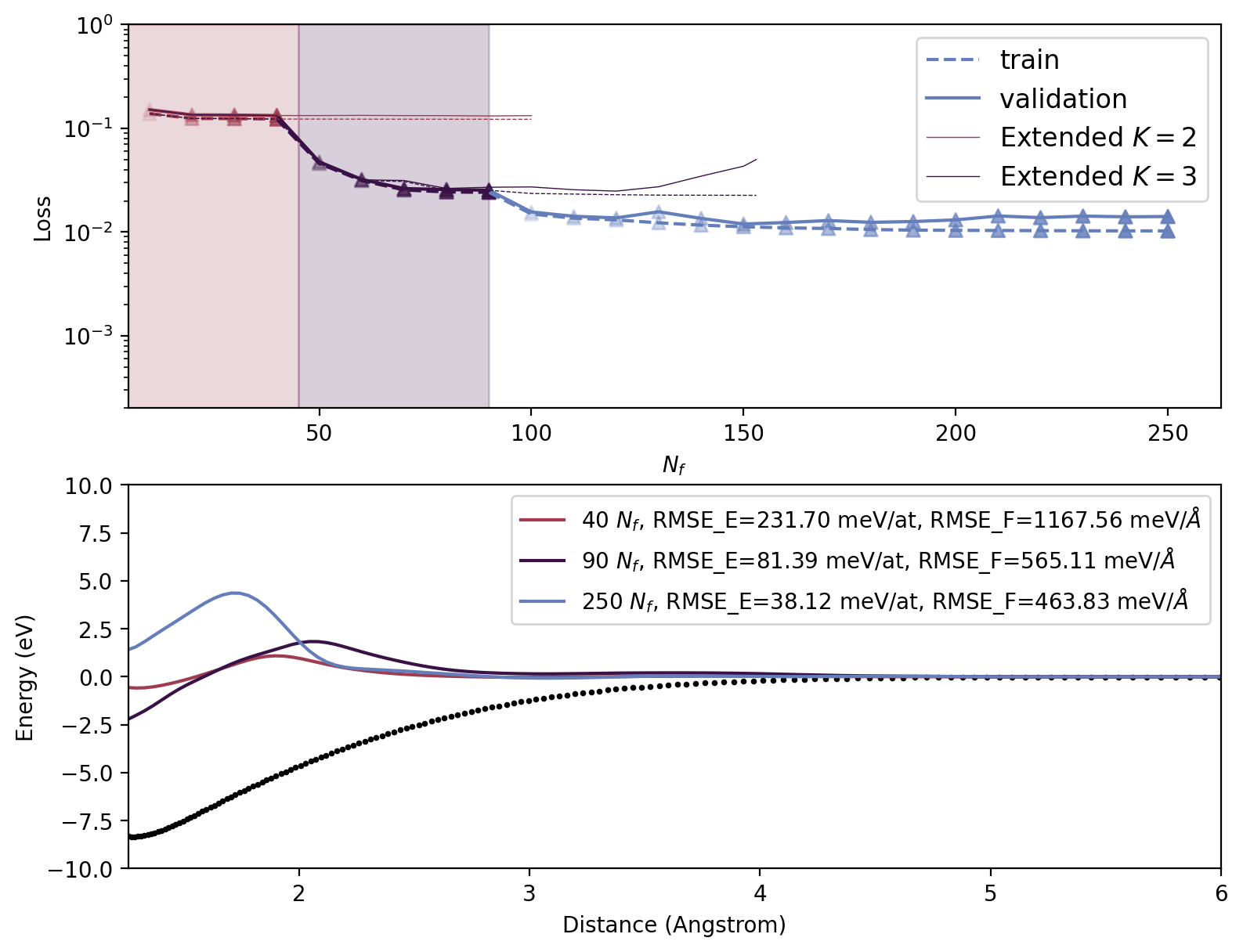}
    \caption{$\mathcal{L}$ vs. $N_f$ (top) and dimer curves (bottom) for the diverse dataset. Orange and pink plots correspond to potentials with only $K=2$ and up to $K=3$ functions, respectively while purple curves include $K=4$ functions. The x-values of the dimer curve start at the minimum distance found in the homogeneous dataset.}
    \label{fig:results_oggen}
\end{figure}

Fig.~\ref{fig:results_oggen} (top) shows the loss $\mathcal{L}$ vs. number of functions $N_f$ for the potential trained on the diverse dataset and (bottom) the corresponding dimer curves at the terminal $N_f$ for each body-order $K$. We see from the loss curves that opening the channel for more complex basis functions by increasing $K$ does help improve training. Further evidence of this comes from the faint lines, which are loss data from potentials still trained with $K=2$ (or $K=3$) but have the same number of functions as the next higher $K$. Despite that, the dimer curve shows that there is a more prominent qualitative change going from $K=2$ to $K=3$ rather than from $K=3$ to $K=4$ despite having a smaller drop in losses. From here we see that the parameters at which the loss converges do not indicate that the curves have already converged.  
\newpage

\subsection{Biasing the diverse dataset with dimer data}
\begin{figure}
    \centering
    \includegraphics[width=0.7\linewidth]{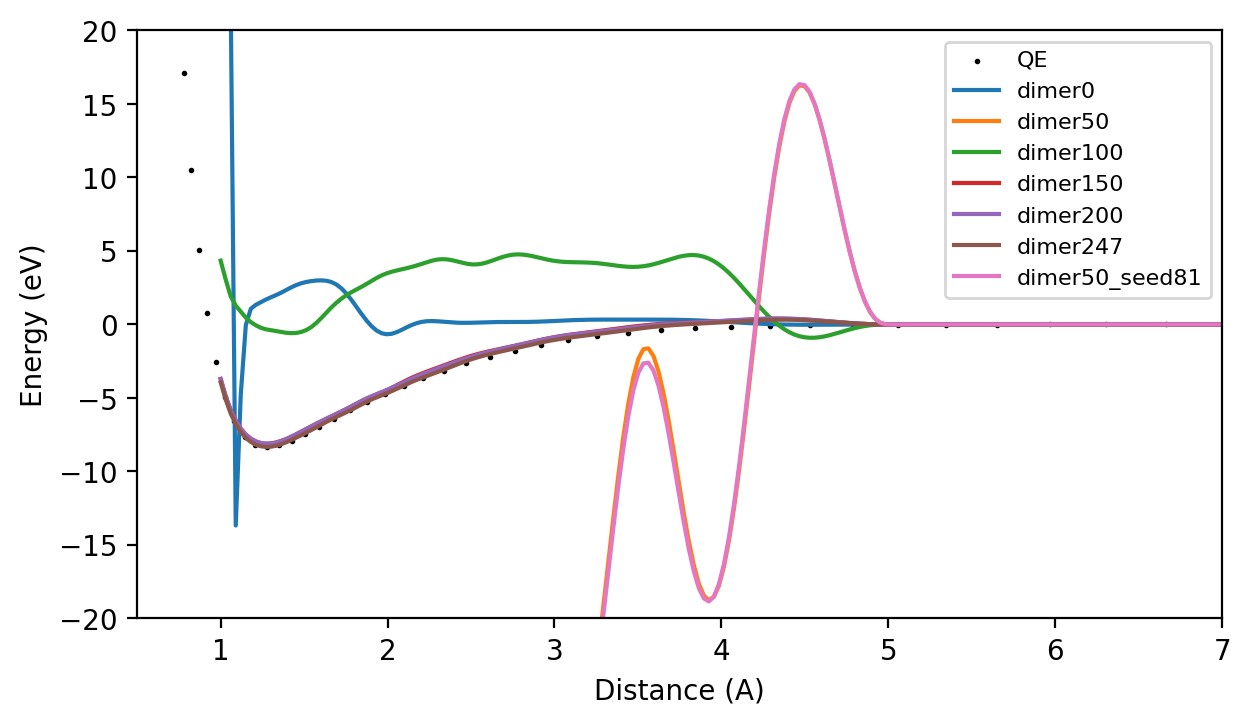}
    \caption{Dimer curves of potentials trained from the mixture of the $1000$ structures from the diverse dataset with an increasing number of structures from the dimer dataset. An additional potential was also trained on the dataset with $50$ dimers added but initialized with a different seed to check the initialization effects. Prominent oscillations at short-range are due to absence of bond pairs in structures in the dataset at the sub-$1$ \AA~distance region.}
    \label{fig:dimgen_dcurve}
\end{figure}

Fig.~\ref{fig:dimgen_dcurve} shows the effect of biasing the diverse dataset with an increasing number of dimer structures for the \verb|pacemaker| tests. These tests were done to verify that the linear ACE potentials is very much capable of learning the correct interaction curve if sufficient data is provided. We show that once around $10\%$ of the dataset is composed of dimer structures, the potential is able to approach the DFT data. We also added an additional potential trained on the dataset with $50$ dimers added but initialized with a different seed to ensure that initialization effects are minimal.

\begin{figure}
    \centering
    \includegraphics[width=0.7\linewidth]{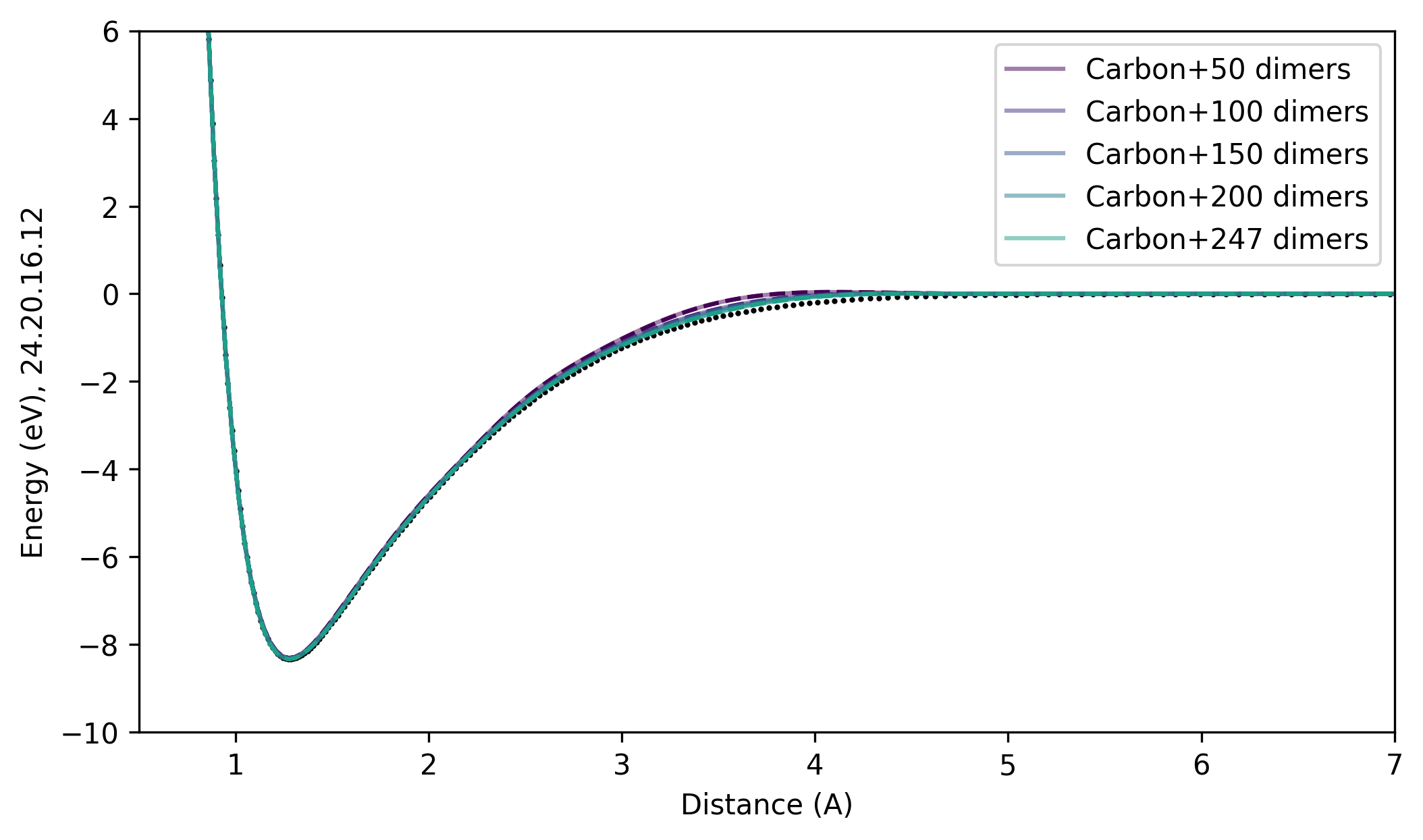}
    \caption{Dimer curves for potentials trained with dimers for carbon. For interpolative tasks we see that linear ACE potentials are able to capture the data well.}
    \label{fig:dimgen_dcurve_acejl}
\end{figure}

Same tests were done for self-interacting and purified potentials for \verb|ACEpotentials.jl| in Fig.~\ref{fig:dimgen_dcurve_acejl}.
\newpage

\subsection{Learning curves of diverse dataset runs}
\begin{figure}
    \centering
    \includegraphics[width=0.7\linewidth]{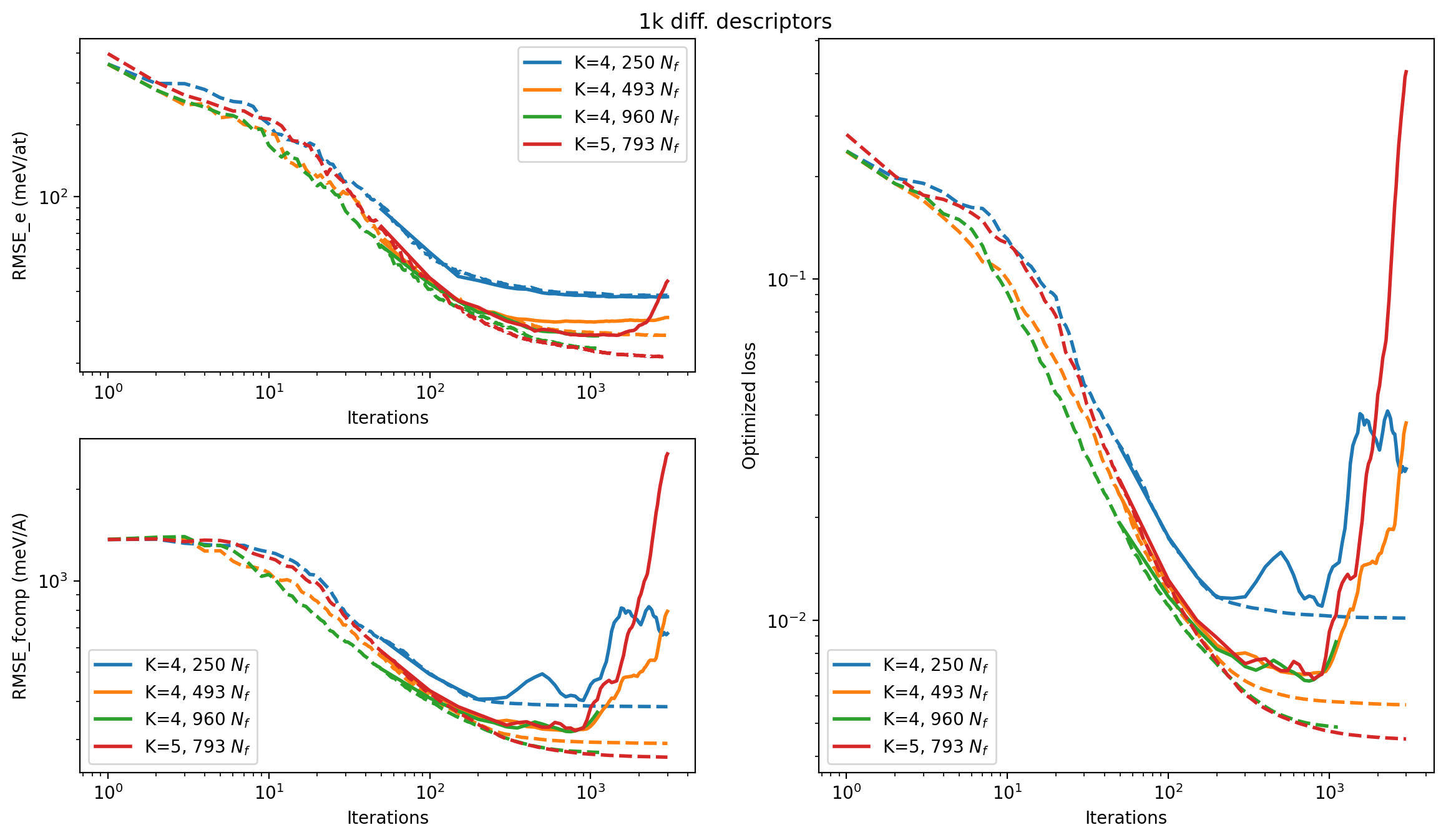}
    \caption{Learning rates for (upper left) energy, (lower left) force, and (right) total for the diverse dataset with $1000$ structures. Different colors indicate different descriptor sets with varying $K$ and $N_f$.}
    \label{fig:learnrate_1k}
\end{figure}

\begin{figure}
    \centering
    \includegraphics[width=0.7\linewidth]{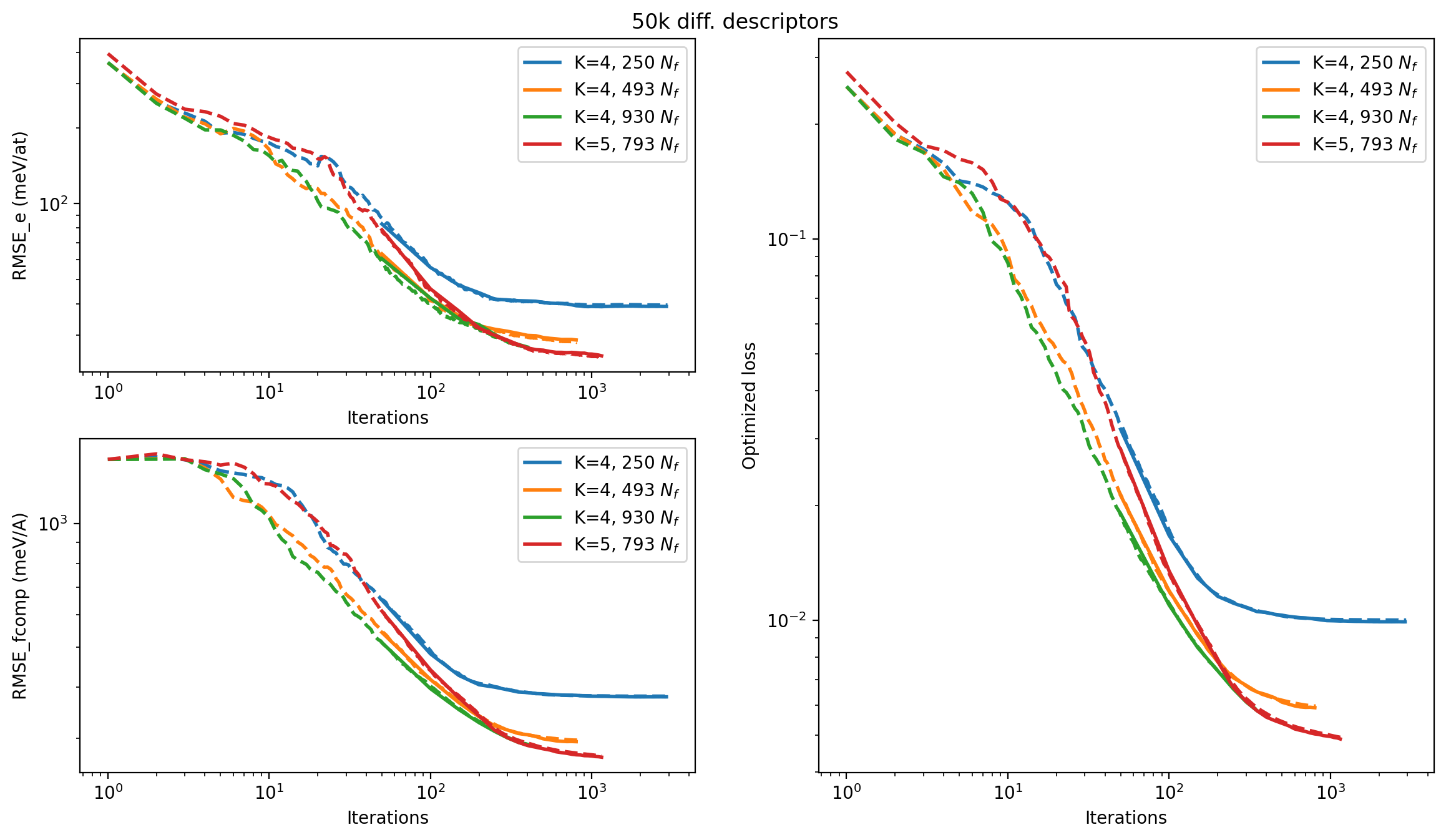}
    \caption{Learning rates for (upper left) energy, (lower left) force, and (right) total for the diverse dataset with $50000$ structures. }
    \label{fig:learnrate_50k}
\end{figure}

Figs.~\ref{fig:learnrate_1k} and ~\ref{fig:learnrate_50k} show the learning rates for the diverse dataset with $1000$ and $50000$ structures, respectively. The different colors indicate the descriptor sets of varying complexity and body-order as controlled by $K$ and $N_f$. In general, we see that the potentials trained by the dataset with fewer structures approach the overfitting regions much faster and indicate that less complex descriptor sets, which have less parameters, are less susceptible to overfitting. On the other hand, the potentials trained with $50$ times the data as compared previously still haven't overfit. Both learning curves have losses around the same order of magnitude, which indicates that the larger dataset benefits from the more complex descriptor sets\footnote{We did not train for longer as it would be computationally expensive without the promise that the losses and dimer curves would have improved by then.}.

\newpage

\subsection{Learning curves and dimer curves for 5k and 10k runs}
\begin{figure}
    \centering
    \includegraphics[width=0.6\linewidth]{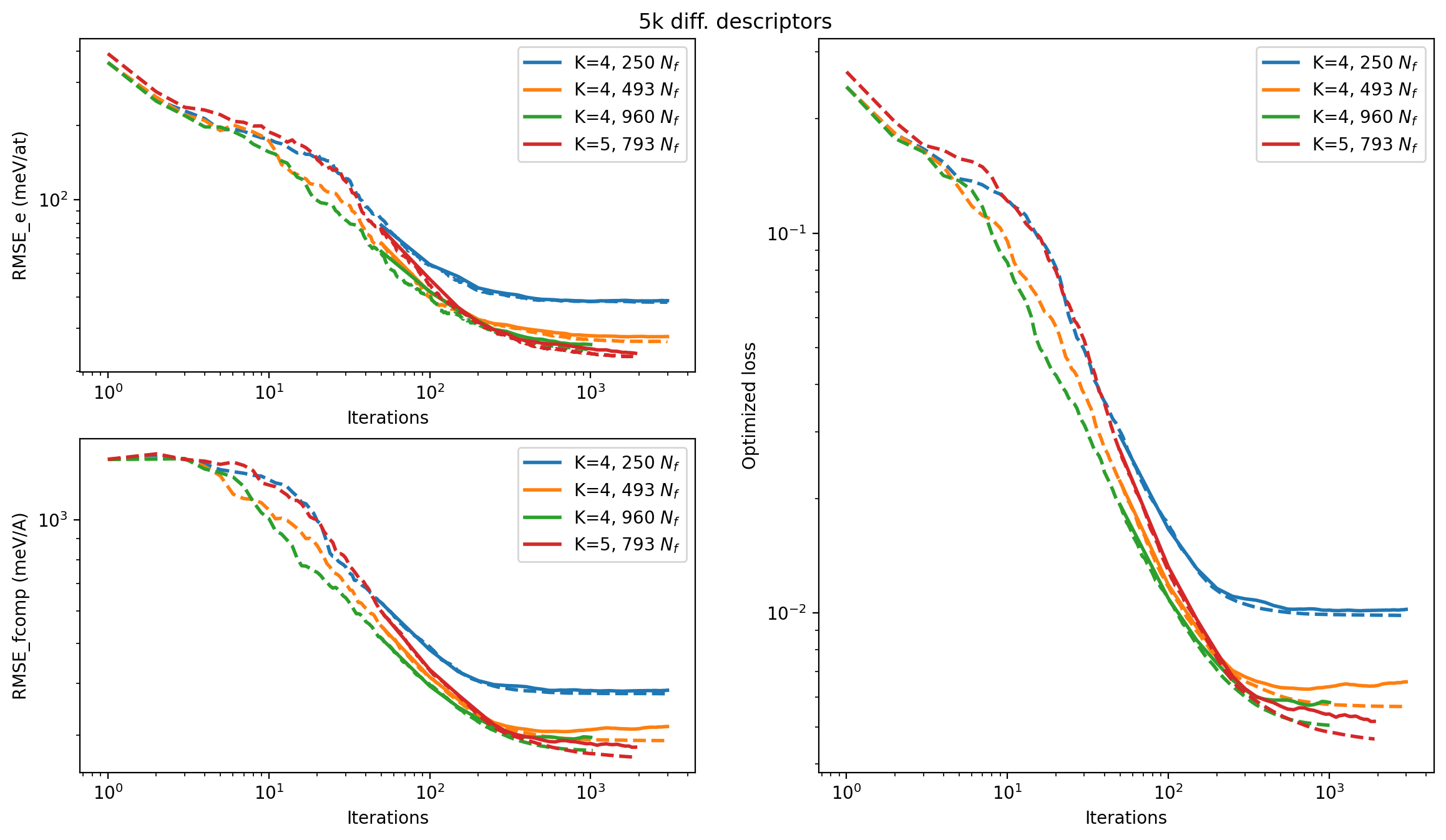}
    \caption{Learning rates for (upper left) energy, (lower left) force, and (right) total for the diverse dataset with $5000$ structures. Different colors indicate different descriptor sets with varying $K$ and $N_f$.}
    \label{fig:learnrate_5k}
\end{figure}

\begin{figure}
    \centering
    \includegraphics[width=0.6\linewidth]{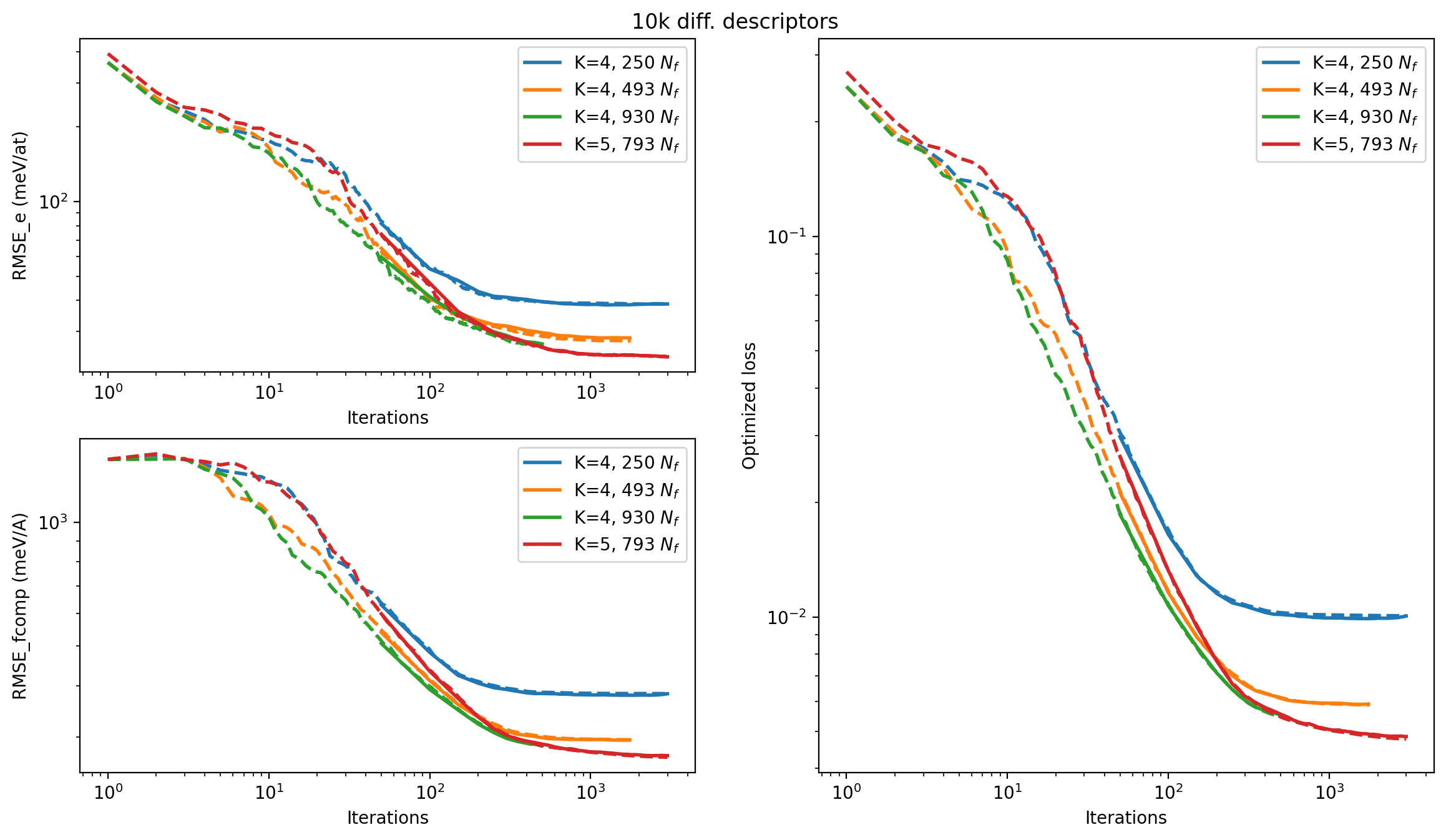}
    \caption{Learning rates for (upper left) energy, (lower left) force, and (right) total for the diverse dataset with $10000$ structures.}
    \label{fig:learnrate_10k}
\end{figure}

\begin{figure}
    \centering
    \includegraphics[width=0.5\linewidth]{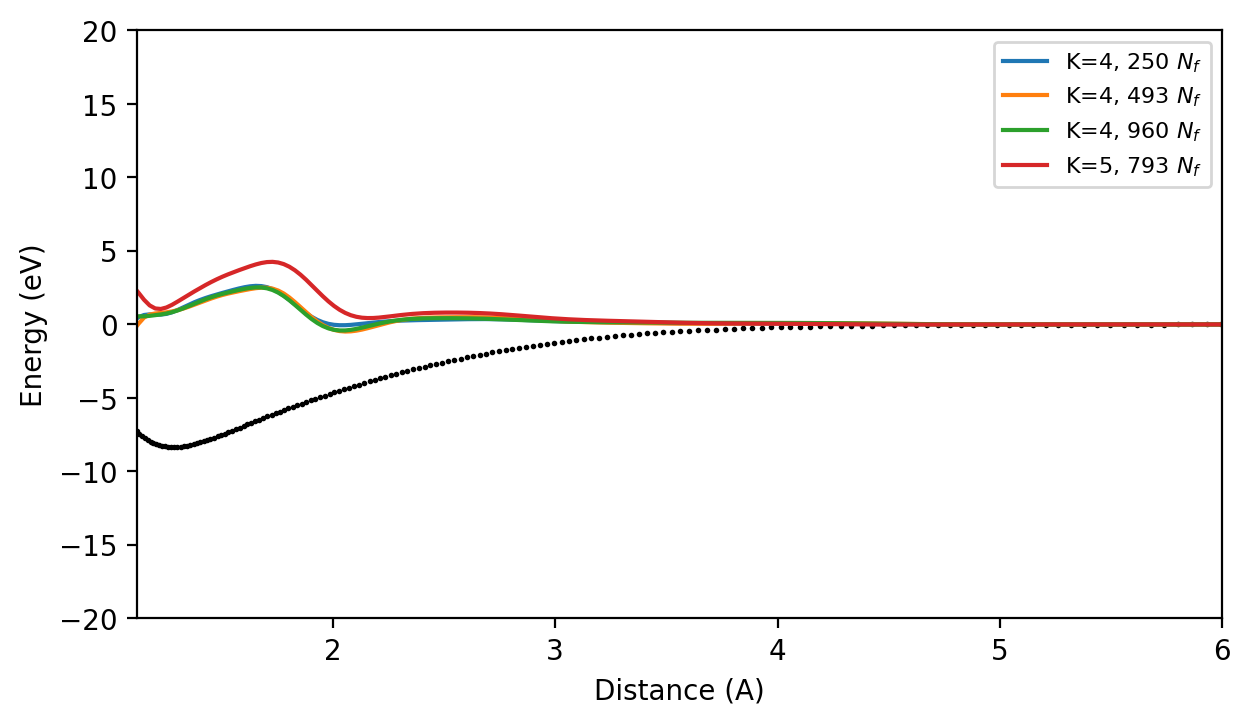}
    \caption{Dimer curves for the diverse dataset with $5000$ structures. Different colors indicate different descriptor sets with varying $K$ and $N_f$.}
    \label{fig:dcurve_5k}
\end{figure}

\begin{figure}
    \centering
    \includegraphics[width=0.5\linewidth]{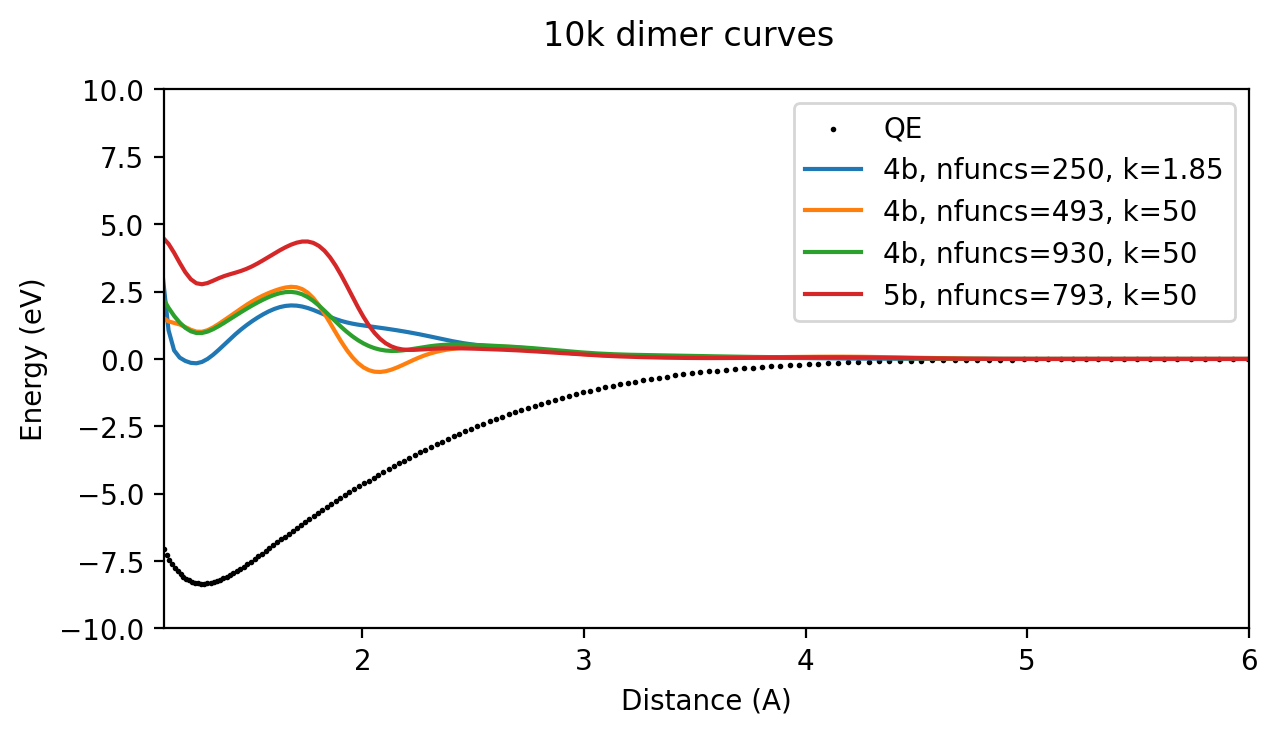}
    \caption{Dimer curves for the diverse dataset with $10000$ structures. Different colors indicate different descriptor sets with varying $K$ and $N_f$.}
    \label{fig:dcurve_10k}
\end{figure}

Figs.~\ref{fig:learnrate_5k} and ~\ref{fig:learnrate_10k} show the learning rates for the diverse dataset with $5000$ and $10000$ structures, respectively. The different colors indicate the descriptor sets of varying complexity and body-order as controlled by $K$ and $N_f$. Less prominent overfitting is observed in the $5000$ dataset as compared to the learning curve of the potentials trained on the $1000$-structure dataset. The learning curves of the $10000$ dataset do not show any signs of overfitting after the runs were completed. Their corresponding dimer curves in Fig.~\ref{fig:dcurve_5k} and ~\ref{fig:dcurve_10k} show similar behavior from the potentials trained from the $1000$ and $50000$ structures in that the dimer curves do not approach a defined curve as higher $K$ basis functions are introduced in the descriptor set.


\subsection{Pair distribution comparison of $1$k and $50$k diverse Carbon datasets}
\begin{figure}
    \centering
    \includegraphics[width=0.7\linewidth]{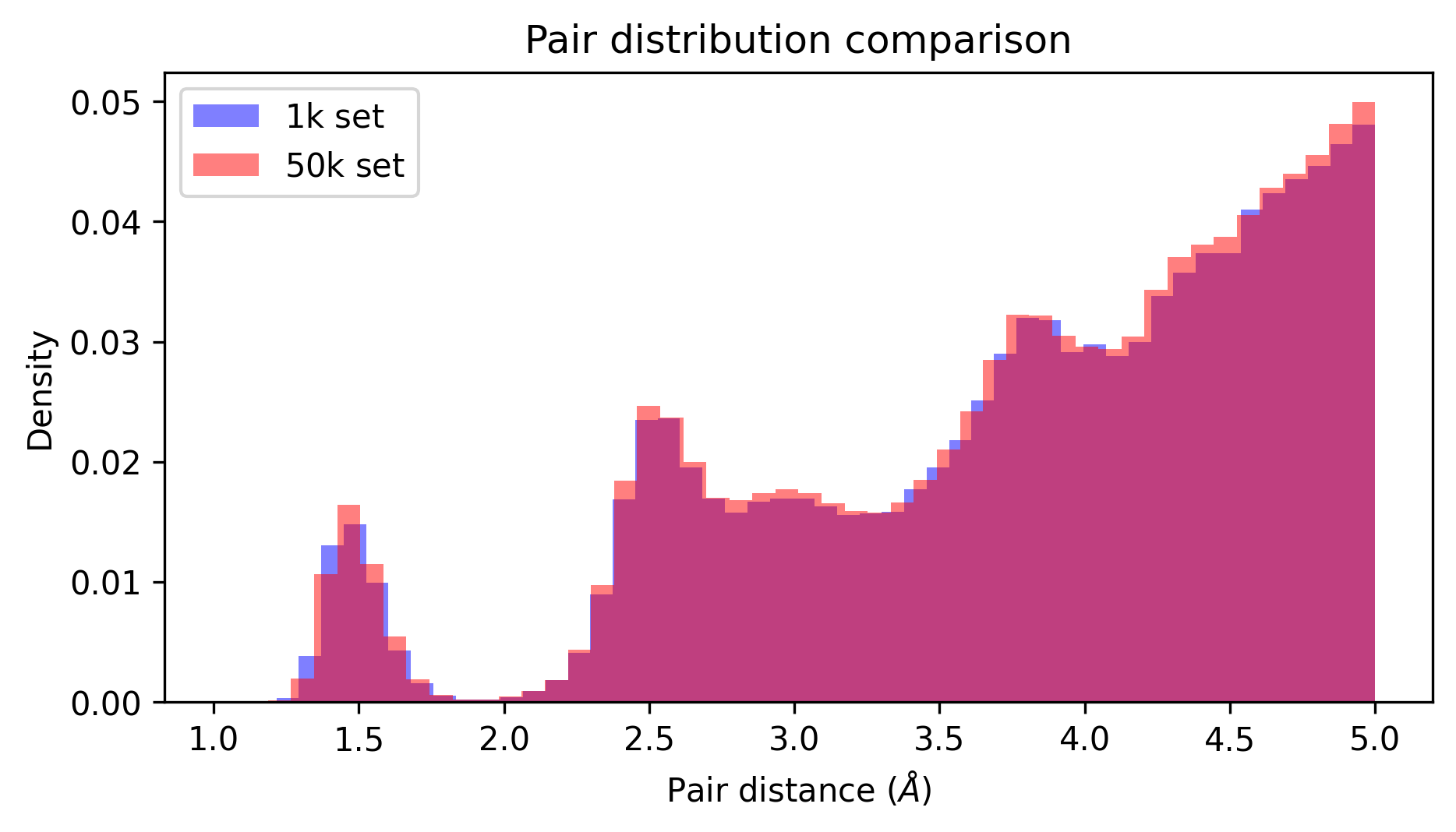}
    \caption{Pair distribution comparisons between the $1000$-structure dataset and the complete $50000$-structure dataset from \citet{shaiduSystematicApproachGenerating2021}.}
    \label{fig:pairdistrib}
\end{figure}

The pair distributions for the $1$k and the $50$k-structure datasets in Fig.~\ref{fig:pairdistrib} are practically identical, with a lack of distances a $2.0$ \AA. The region may be supplemented by more varied structures ($n$-mers, high-pressure structures) as done in \citet{qamarAtomicClusterExpansion2023}, but a separate investigation must be done for this dataset.

\section{Model weights and average features of in-domain tests}

The mean absolute model weights, $c$, with respect to the features corresponding to each body-order $K$ are tabulated in Table~\ref{tab:modelweights_pure} and~\ref{tab:modelweights_si} for the purified and self-interacting bases, respectively. These results verify that potentials with purified bases have negligible coefficients for features $K$ that have orders higher than the maximum number of atoms in the structures of the training set. Potentials with the original ACE bases will have higher order $K$ features that are relevant due to the spurious self-interactions masked as lower-order terms. The peculiar case of the Si$_2$-trained potentials stem from the fact that the $K=2$ functions used ($24$ functions in the chosen basis) are already sufficient in capturing the pair interactions of the training set which only contains dimers. This applies to both purified and self-interacting bases, making the self-interacting basis functions practically unnecessary in capturing the correct two-body energetics.

\begin{table}[h]
\caption{Mean absolute model weights for the features each $K$ of the in-domain tests for the purified bases}
\label{tab:modelweights_pure}
\begin{tabular}{|l|l|rrrr|}
\hline
\multirow{2}{*}{Trainset} & \multirow{2}{*}{Basis} & \multicolumn{4}{c|}{Feature} \\ \cline{3-6} 
 &  & \multicolumn{1}{r|}{$K=2$} & \multicolumn{1}{r|}{$K=3$} & \multicolumn{1}{r|}{$K=4$} & $K=5$ \\ \hline
\multirow{3}{*}{Si$_2$} & 24.20 & \multicolumn{1}{r|}{$2.024$} & \multicolumn{1}{r|}{$<10^{-8}$} & \multicolumn{1}{r|}{} &  \\ \cline{2-6} 
 & 24.20.16 & \multicolumn{1}{r|}{$2.024$} & \multicolumn{1}{r|}{$<10^{-8}$} & \multicolumn{1}{r|}{$<10^{-8}$} &  \\ \cline{2-6} 
 & 24.20.16.12 & \multicolumn{1}{r|}{$2.024$} & \multicolumn{1}{r|}{$<10^{-8}$} & \multicolumn{1}{r|}{$<10^{-8}$} & $<10^{-8}$ \\ \hline
\multirow{3}{*}{Si$_{23}$} & 24.20 & \multicolumn{1}{r|}{$1.85$} & \multicolumn{1}{r|}{$0.392$} & \multicolumn{1}{r|}{} &  \\ \cline{2-6} 
 & 24.20.16 & \multicolumn{1}{r|}{$1.85$} & \multicolumn{1}{r|}{$0.392$} & \multicolumn{1}{r|}{$<10^{-8}$} &  \\ \cline{2-6} 
 & 24.20.16.12 & \multicolumn{1}{r|}{$1.85$} & \multicolumn{1}{r|}{$0.392$} & \multicolumn{1}{r|}{$<10^{-8}$} & $<10^{-8}$ \\ \hline
\multirow{3}{*}{Si$_{234}$} & 24.20 & \multicolumn{1}{r|}{$1.569$} & \multicolumn{1}{r|}{$0.421$} & \multicolumn{1}{r|}{} &  \\ \cline{2-6} 
 & 24.20.16 & \multicolumn{1}{r|}{$1.716$} & \multicolumn{1}{r|}{$0.296$} & \multicolumn{1}{r|}{$0.495$} &  \\ \cline{2-6} 
 & 24.20.16.12 & \multicolumn{1}{r|}{$1.716$} & \multicolumn{1}{r|}{$0.296$} & \multicolumn{1}{r|}{$0.495$} & $<10^{-8}$ \\ \hline
\multirow{3}{*}{Si$_{2345}$} & 24.20 & \multicolumn{1}{r|}{$1.429$} & \multicolumn{1}{r|}{$0.315$} & \multicolumn{1}{r|}{} &  \\ \cline{2-6} 
 & 24.20.16 & \multicolumn{1}{r|}{$1.420$} & \multicolumn{1}{r|}{$0.291$} & \multicolumn{1}{r|}{$0.434$} &  \\ \cline{2-6} 
 & 24.20.16.12 & \multicolumn{1}{r|}{$1.370$} & \multicolumn{1}{r|}{$0.314$} & \multicolumn{1}{r|}{$0.482$} & $0.574$ \\ \hline
\end{tabular}
\end{table}

\begin{table}[h]
\caption{Mean absolute model weights for the features each $K$ of the in-domain tests for the self-interacting bases}
\label{tab:modelweights_si}
\begin{tabular}{|l|l|rrrr|}
\hline
\multirow{2}{*}{Trainset} & \multirow{2}{*}{Basis} & \multicolumn{4}{c|}{Feature} \\ \cline{3-6} 
 &  & \multicolumn{1}{r|}{$K=2$} & \multicolumn{1}{r|}{$K=3$} & \multicolumn{1}{r|}{$K=4$} & $K=5$ \\ \hline
\multirow{3}{*}{Si$_2$} & 24.20 & \multicolumn{1}{r|}{$2.024$} & \multicolumn{1}{r|}{$<10^{-8}$} & \multicolumn{1}{r|}{} &  \\ \cline{2-6} 
 & 24.20.16 & \multicolumn{1}{r|}{$2.024$} & \multicolumn{1}{r|}{$<10^{-8}$} & \multicolumn{1}{r|}{$<10^{-8}$} &  \\ \cline{2-6} 
 & 24.20.16.12 & \multicolumn{1}{r|}{$2.024$} & \multicolumn{1}{r|}{$<10^{-8}$} & \multicolumn{1}{r|}{$<10^{-8}$} & \multicolumn{1}{r|}{$<10^{-8}$} \\ \hline
\multirow{3}{*}{Si$_{23}$} & 24.20 & \multicolumn{1}{r|}{$1.85$} & \multicolumn{1}{r|}{$0.392$} & \multicolumn{1}{r|}{} &  \\ \cline{2-6} 
 & 24.20.16 & \multicolumn{1}{r|}{$1.84$} & \multicolumn{1}{r|}{$0.330$} & \multicolumn{1}{r|}{$0.146$} &  \\ \cline{2-6} 
 & 24.20.16.12 & \multicolumn{1}{r|}{$1.85$} & \multicolumn{1}{r|}{$0.329$} & \multicolumn{1}{r|}{$0.142$} & $0.0779$ \\ \hline
\multirow{3}{*}{Si$_{234}$} & 24.20 & \multicolumn{1}{r|}{$1.569$} & \multicolumn{1}{r|}{$0.421$} & \multicolumn{1}{r|}{} &  \\ \cline{2-6} 
 & 24.20.16 & \multicolumn{1}{r|}{$1.714$} & \multicolumn{1}{r|}{$0.359$} & \multicolumn{1}{r|}{$0.485$} &  \\ \cline{2-6} 
 & 24.20.16.12 & \multicolumn{1}{r|}{$1.708$} & \multicolumn{1}{r|}{$0.333$} & \multicolumn{1}{r|}{$0.431$} & $0.315$ \\ \hline
\multirow{3}{*}{Si$_{2345}$} & 24.20 & \multicolumn{1}{r|}{$1.429$} & \multicolumn{1}{r|}{$0.315$} & \multicolumn{1}{r|}{} &  \\ \cline{2-6} 
 & 24.20.16 & \multicolumn{1}{r|}{$1.410$} & \multicolumn{1}{r|}{$0.280$} & \multicolumn{1}{r|}{$0.372$} &  \\ \cline{2-6} 
 & 24.20.16.12 & \multicolumn{1}{r|}{$1.355$} & \multicolumn{1}{r|}{$0.371$} & \multicolumn{1}{r|}{$0.497$} & $0.521$ \\ \hline
\end{tabular}
\end{table}

\bibliography{y2ace}

\end{document}